\documentclass[a4paper,aps,onecolumn,nofootinbib,preprintnumbers,superscriptaddress,amsmath,amssymb,amsfonts]{revtex4}
\usepackage{graphicx}
\usepackage{bm,natbib}
\usepackage{amsmath}
\usepackage[english]{babel}
\usepackage[T1]{fontenc}
\usepackage{amssymb}
\usepackage{amsfonts}
\usepackage{epsfig}
\usepackage{colordvi}
\usepackage{psfrag}
\usepackage{color}
\usepackage{ mathrsfs }

\newcommand{\G}{\mathcal{G}}

\newcommand{\La}{\mathscr{L}}
\usepackage{dcolumn}
\usepackage{multirow}
\usepackage{hyperref}
\usepackage{epstopdf}
\usepackage{bm}
\usepackage{rotating}
\usepackage{lscape}
\usepackage{times}
\usepackage{comment}

\usepackage{tikz,xcolor}

\definecolor{lime}{HTML}{A6CE39}
\DeclareRobustCommand{\orcidicon}{
	\begin{tikzpicture}
	\draw[lime, fill=lime] (0,0) 
	circle [radius=0.16] 
	node[white] {{\fontfamily{qag}\selectfont \tiny ID}};
	\draw[white, fill=white] (-0.0625,0.095) 
	circle [radius=0.007];
	\end{tikzpicture}
	\hspace{-2mm}
}

\foreach \x in {A, ..., Z}{\expandafter\xdef\csname orcid\x\endcsname{\noexpand\href{https://orcid.org/\csname orcidauthor\x\endcsname}
			{\noexpand\orcidicon}}
}

\hypersetup{
  colorlinks=true,        % false: boxed links; true: colored links
  linkcolor=blue,         % color of internal links
  citecolor=magenta,      % color of links to bibliography
}

\begin{document}
\title{DNA Mutations via Chern-Simons Currents}

\author{Francesco Bajardi\orcidA{}}
%\email{francesco.bajardi@unina.it} 
\affiliation{Dipartimento di Fisica "Ettore Pancini", Universit\`{a} degli Studi di Napoli
"Federico II", Compl. Univ. di Monte S. Angelo, Edificio G, Via Cinthia, I-80126, Napoli, Italy}
\affiliation{INFN Sezione  di Napoli, Compl. Univ. di Monte S. Angelo, Edificio G, Via Cinthia, I-80126, Napoli, Italy}
%\item \href{https://orcid.org/0000-0000-0000-0000}{\textcolor{orcidlogocol}{\aiOrcid} \hspace{2mm} orcid.org/0000-0000-0000-0000}

\author{Lucia Altucci\orcidB{}}
\email{lucia.altucci@unicampania.it}
\affiliation{Dipartimento di Medicina di Precisione, Universit\`{a} degli Studi della Campania "L. Vanvitelli", Napoli, Italy.}
\affiliation{Biogem "Istituto di Biologia molecolare e genetica", 83031 Ariano Irpino, Italy}

\author{Rosaria Benedetti\orcidC{}}
%\email{rosaria.benedetti@unicampania.it}
\affiliation{Dipartimento di Medicina di Precisione, Universit\`{a} degli Studi della Campania "L. Vanvitelli", Napoli, Italy.}

\author{Salvatore Capozziello\orcidD{}}
\email{capozziello@na.infn.it}
\affiliation{Dipartimento di Fisica "Ettore Pancini", Universit\`{a} degli Studi di Napoli
"Federico II", Compl. Univ. di Monte S. Angelo, Edificio G, Via Cinthia, I-80126, Napoli, Italy}
\affiliation{INFN Sezione  di Napoli, Compl. Univ. di Monte S. Angelo, Edificio G, Via Cinthia, I-80126, Napoli, Italy}
\affiliation{Scuola Superiore Meridionale, Largo San Marcellino 10, 80138 Napoli, Italy}

\author{Maria Rosaria Del Sorbo\orcidE{}}
%\email{Mariarosaria.delsorbo@unina.it}
\affiliation{Istituto Statale d'Istruzione Superiore "Leonardo da Vinci", via F. Turati Poggiomarino, Naples, Italy}
\affiliation{Dipartimento di Ingegneria Industriale, Universit\`{a} degli Studi di Napoli
"Federico II", Via Claudio n.21, I-80125, Napoli, Italy}

\author{Gianluigi Franci\orcidF{}}
%\email{gfranci@unisa.it}
\affiliation{Department of Medicine, Surgery and Dentistry "Scuola Medica Salernitana", University of Salerno, 84081 Baronissi SA, Italy}
\affiliation{AOU "San Giovanni di Dio e Ruggi d'Aragona", Salerno}

\author{Carlo Altucci\orcidG{}}
\email{carlo.altucci@unina.it}
\affiliation{INFN Sezione  di Napoli, Compl. Univ. di
Monte S. Angelo, Edificio G, Via Cinthia, I-80126, Napoli, Italy}
\affiliation{Dipartimento di Scienze Biomediche Avanzate, Universit\`{a} degli Studi di
Napoli "Federico II", via Pansini 5, Napoli, Italy}

\date{\today}

\begin{abstract}
We test the validity of a possible schematization of DNA structure and dynamics based on the Chern-Simons theory, that is a topological field theory mostly considered in the context of effective gravity theories. By means of the expectation value of the Wilson Loop, derived from this analogue gravity approach, we find the point-like curvature of genomic strings in KRAS human gene and COVID-19 sequences, correlating this curvature with the genetic mutations. The point-like curvature profile, obtained by means of the Chern-Simons currents, can be used to infer the position of the given mutations within the genetic string. Generally, mutations take place in the highest Chern-Simons current gradient locations and subsequent mutated sequences appear to have a smoother curvature than the initial ones, in agreement with a free energy minimization argument. 
\end{abstract}

\maketitle

\section{Introduction} 
Genomic strings schematization methods represent one of the most controversial and discussed branch of science. In this scenario, the application of those methods to DNA alignment is still not fully uncovered. Several approaches aim to exhaustively predict the evolution of macro molecules, in order to get information regarding their spatial configuration \cite{Baldi, Waterman, Gusfield, Vinga}. However, a complete theory capable of predicting the interactions occuring among macro molecules and the corresponding biological implications is still missing. Biological systems, such as nucleic acids or protein, often exhibit complicated topological structures, since several parts of the same molecule may assume a non-trivial three-dimensional shape, called tertiary structure. When two or more tertiary structures interact, the resulting system fold into a quaternary structure. In this framework, schematization approaches are particularly important in view of understanding the spatial configuration assumed by the system and, consequently, the interactions occurring among neighboring elements which may be located hundreds of kilobases away from each other and, in some cases, also in different chromosomes \cite{PMID}.

As an example, from the spatial configuration assumed by the DNA, it is possible to infer the place in which genomic mutations might occur, as well as the consequent difference among phenotypes. Schematization approaches can also help to provide the genetic (and epigenetic) probability to develop a certain disease. Another example is given by the interaction between proteins and virus genome which, if well described, can lead to a comprehension of the corresponding infection evolution. Standard modeling techniques are mostly based on probability considerations, aimed at outlining the many body interactions by means of statistical mechanics \cite{Lengauer, Morris, Kahraman}.

In this paper we want to test an innovative method for the schematization of biomolecule configurations, based on the topological Chern-Simons theory. It mainly relies on the curvature assumed by biological systems, using the numerical value of the Chern-Simons current, namely the expectation value of the Wilson loop \cite{Maldacena:1998im, Lepage:1992xa, Aoki:2006br, Appelquist:1981vg}. 

Indeed, from very general and basic theories such as classical and quantum theories of gravity, ideas can lead to far beyond closely related fields, such as theoetical physics, cosmology and astrophysics, to push concepts and applications to complex systems, there including the interactions between biomolecules, such as nucleic acids and proteins. This model is significant because it introduces a new approach to treat biological systems, which differs from standard bioinformatics methods as it is not based on approaches typical of statistical mechanics applied to complex systems, but rather on first principles of field theories of physics. This novel point of view might be used completing outputs derived from statistical methods, to address issues of biological and medical sciences, such as preventing diseases, predicting the evolution of a genetic string or investigating the docking among biological large molecules, potentially implementing the nowadays knowledge of the biological scenario. The link between gravitational theories and the dynamics/interactions of complex biomolecules is the topological nature of the former which can be essential to describe the complicated physical-chemical and biological behavior of the latter, very much relying on their topology. Basically, the main idea is to describe the DNA curvature by using the same formalism used for the space-time, treating the interactions occurring in biological systems as driven by the same general principles that govern the gravitational interaction.

Moreover, the deterministic approach based on Chern-Simons gravity can be also merged with the intrinsic probabilistic aspect of standard bioinformatic techniques in different ways. As an example, using topological field theories to describe DNA configuration can provide the exact position in which mutations take place, by means of the comparison between two sequence curvatures. Once the position of the mutation is identified, bioinformatics is able to predict the probabilistic evolution and the clinical impact of that mutation. Another potential application which can be considered in the context of Chern-Simons formalism, is the docking between macro-molecules \cite{Capozziello:2018lnd}. The latter can be understood as the interaction among different points, which tend to attract each other only where the corresponding curvatures are similar (by analogy with the gravitational interaction). Also in this regard, the probabilistic vision provided by bioinformatic techniques can be combined with the prediction given by topological field theories, in order to develop a coherent scheme capable of predicting where and when a disease could manifest. 
 
Although the application of Chern-Simons gravity to complex systems seems to be unusual, topological field theories are deeply studied in several branches of physics, due to their suitability at ultraviolet (UV) and infrared (IR) scales \cite{Zanelli:2005sa, Aharony:2008ug, Witten:1988hf, Son:2002sd, Achucarro:1987vz}. In general, they involve \emph{Topological Invariants}, namely quantities which are conserved under homeomorphism transformations. Topological invariants, indeed, only depend on the spacetime topology, independently of the point-like geometry \cite{Nakahara:2003nw}. They find their best application in the description of the gravitational interaction, and are considered to the purpose of finding alternatives to General Relativity (GR) which better adapt to the quantum formalism \cite{Nojiri:2005jg, Nojiri:2005vv, Wheeler:1985nh, Buchel:2009sk, Li:2007jm, Chamseddine:1989nu, Bajardi:2020xfj, Bajardi:2020osh, Bajardi:2019zzs}. 

Moreover, although the theoretical predictions of GR are perfectly consistent with observations at the level of solar system, the theory suffers some shortcoming at larger scales. As an example, the late-time accelerated expansion of the universe is nowadays addressed to a never detected form of energy, called Dark Energy. Similarly,  incompatibilities in the galaxy rotation curve led to the introduction of Dark Matter, which is supposed to account for the 85$\%$ of matter in the Universe and to have had a high influence in the evolution of the latter. These are two of the biggest problems suffered by GR; for a complete discussion see \emph{e.g.} \cite{Padmanabhan:2007xy, Barack:2018yly, Bull:2015stt, Capozziello:2011et, Capozziello:2019klx}

With the aim to solve part of these issues, mainly those related to a self-consistent quantization of the gravitational interaction, in the first half of twentieth century, S.S. Chern and J.H. Simons developed a topological field theory capable of describing gravity as a gauge invariant theory of different gauge groups \cite{Chern:1974ft}. It turns out that n-dimensional Lagrangians whose exterior derivative gives $n+1$-dimensional topological invariants, are \emph{quasi}-gauge invariant, \emph{i.e} they only change by a surface term after performing a gauge transformation. However, the lack of non-trivial topological invariants in even dimensions, restricts the validity of the formalism to odd dimensions, only. This is the main obstacle toward the construction of a 3+1 dimensional topological theory of gravity, though odd-dimensional topological theories find large applications in several fields. See \emph{e.g.} \cite{Zanelli:2005sa, Aharony:2008ug, Witten:1988hf, Son:2002sd, Achucarro:1987vz} for basic foundations of Chern-Simons gravity and \cite{Cai:1998vy, Smolin:1994qb, Giombi:2011kc, Susskind:2001fb} for applications. 

Due to the applications to the three-dimensional electromagnetic theory \cite{Carlip:2008jk, Jackiw:1998js}, one of the most studied Chern-Simons Lagrangian is the 2+1 dimensional $U(1$)-invariant Lagrangian, namely: 
\begin{equation}
\La^{(3)}_{CS} = \textbf{A} \text{d}\textbf{A},
\label{LACS3}
\end{equation}
with $\textbf{A}$ being the one-form connection and $\text{d}\textbf{A}$ its exterior derivative. Notice that the exterior derivative of $\La^{(3)}_{CS}$ provides the four-dimensional Pontryagin density, namely $P^{(4)} = F \wedge F$, where $F$ represents the two-form curvature defined as $F =\text{d}\textbf{A}$. 

The Lagrangian in Eq. \eqref{LACS3} can be also applied to standard electromagnetic theory, providing a massive wave equation which carries extra polarization modes. Specifically, in coordinates representation, the Chern-Simons term can be considered along with the free electromagnetic Lagrangian to provide a massive wave equation of the form $(\Box + m^2) \epsilon^{\mu \nu \rho} F_{\mu \nu} = 0$, with $m$ being a constant having mass dimension, $\Box$ the d'Alembert operator $\Box \equiv \partial_\mu \partial^\mu$, $F_{\mu \nu}$ the electromagnetic tensor and $ \epsilon^{\mu \nu \rho}$ the Levi-Civita symbol. 

Another well studied Lagrangian is the $SU(N)$-invariant three-dimensional Chern-Simons Lagrangian
\begin{equation}
\La^{(3)}_{CS} = \text{tr} \left[\textbf{A} \text{d}\textbf{A} + \frac{2}{3}\textbf{A} \wedge \textbf{A} \wedge \textbf{A} \right],
\label{SUNLagr}
\end{equation}
whose exterior derivative yields the $SU(4)$-invariant Pontryagin density $P^{(4)} = \text{tr} \left[F \wedge F\right]$. Lagrangian in Eq. \eqref{SUNLagr} is mostly studied due to the applications to supergravity and string theory \cite{Achucarro:1987vz, Witten:1992fb, Ooguri:1999bv}. 

Another topological invariant, used to construct a gauge-invariant Lagrangian, is the four-dimensional Euler density, which turns out to be the exterior derivative of the three-dimensional Anti de Sitter-invariant Chern--Simons Lagrangian, that is:
\begin{equation}
\La_3^{AdS} = \epsilon_{\mu \nu \rho} \left( R^{\mu \nu} \wedge e^\rho + \frac{1}{3l^2} e^\mu \wedge e^\nu \wedge e^\rho \right)\,,
\end{equation}
with $ \epsilon_{\mu \nu \rho}$ being the Levi-Civita symbol, $R^{\mu \nu}$ the two-form curvature, $\wedge$ the exterior product and $l$ a real constant with dimension of length. In general, the $2n-1$-dimensional Chern-Simons Lagrangian, invariant under the local Anti de Sitter group, reads:
\begin{equation}
\La^{AdS}_{2n-1} = \sum_{i=0}^{n-1} \tilde{\alpha}_i \La^{(2n-1,i)}\,, \qquad \;\;\; \tilde{\alpha}_i = \frac{(\pm 1)^{i+1} l^{2i-n}}{n-2i} \binom{n-1}{i}\,.
\label{AdS2D-1}
\end{equation}
Due to the AdS/CFT\footnote{Anti-de Sitter/Conformal Field Theory} correspondence, the above Lagrangian is mainly considered in five dimensions, where cosmological and spherically symmetric solutions can be analytically found \cite{Chapline:1982ww, Benna:2008zy, Chamseddine:1990gk, Aviles:2016hnm, Gomez:2011zzd, Miskovic:2007mg, Miskovic:2010ui, Kofinas:2008ub, Cardoso:2009pk, Molina:2010fb, Bajardi:2021hya}. 

The Chern-Simons approach, as we are going to discuss, can represent a starting point for the analysis of biological systems. From a conceptual point of view, the issue comes out because Quantum Mechanics, being a linear theory, could not be sufficient to approach the high non-linearity of biological systems. Due to this, the latter can be suitably described by non-linear theories like GR or Chern-Simons.

For instance, by means of the Chern-Simons formalism, some biological  problems can be addressed, such as the presence of knotted DNAs and their interactions with proteins \cite{Dabrowski}. Furthermore, in \cite{Lachner} the interactions of unknotted RNAs with knotted proteins have been analyzed in the process of codon and correction of RNA in methil transfer, as well as a general equation to solve the dynamics of knotted proteins has been proposed by Lin and Zewail \cite{Zewail}, based on the Wilson loop operator for gene expression with a boundary phase condition. 

On the other hand, basic foundations lying behind the application of Chern-Simons theory to biology can be found in \cite{Capozziello:2017zfk} and  \cite{Capozziello:2018lnd}. In these references, the authors develop the formal structure of the theory and consider some application to biological system, in order to unveil the mechanism of DNA-RNA transcriptions. They also provide some insights to specifically describe the junk area within the DNA sequence \cite{Capozziello:2017zfk}. In \cite{Capozziello:2017zfk}, the theory is applied to  the docking mechanism of biological macro-molecules, such as the configurational dynamics occurring in protein-protein interactions. 

Without claiming completeness, in Sec. \ref{sect2} we outline the main properties of the theory, with the aim to subsequently test its validity by considering DNA sequences and introducing known mutations. The introduction of a mutation yields a change in the point-like curvature of the given sequence, which may give important information regarding the biological impact carried by such mutation. From the mutated sequence, it is possible to infer the frequency/probability of the mutation to occur, as well as to predict the evolution of the system towards a given configuration. 

This paper is organized as follows: in Sec. \ref{sect2} we briefly review the application of Chern-Simons theory to DNA and RNA systems; in Sec. \ref{sect3} the formalism is then applied to different strings of KRAS human gene and to SARS-CoV-2 virus sequences. In the former case, we apply the model to analyze mutations occurring in few regions of the KRAS human gene. The latter is a gene  acting as an on/off switch in cell signaling which, among its functions, controls cell proliferation. When KRAS is mutated, negative signaling is disrupted, with the consequence that cells can continuously proliferate, often degenerating into tumors \cite{hartman, Kronenburg}. 

In our analysis KRAS sequences with mutations are thus compared with reference sequences, with the aim to use Chern-Simons theory to infer predictions of biological interest. As for the latter case, which is naturally one of the most studied RNA sequence to date due to pandemic, using a genome wide approach, Bobay \emph{et al.} \cite{LMBOBAY} examined SARS-CoV-2 RNA, observing that recombination events account for approximately 40\% of the polymorphisms, and gene exchange occurs only within strains of the same subgenus (Sarbeco virus). Moreover, frequent mutations tend to increase the likelihood of convergent mutations, in regions exposed to a major positive selection, causing analogies in the sequences that could be misinterpreted as it was a recombination, and introduce new diversifying mutations which might accumulate, hiding past recombination events \cite{LMBOBAY}.

Genomic sequences of various SARS-CoV-2 strains from all over the world are available on specific platforms (eg. \href{https://www.gisaid.org/}{GISAID}) and increasingly monitored to timely track SARS-CoV-2 variants \cite{MRISLAM}; as large databases and systematic sequencing are required, irregular sampling in time and space represents a crucial limitation to track pandemic evolution. Genetic diversity observed in SARS-CoV-2 populations across distinct geographic areas suggests independent events of SARS-CoV-2 introduction occurred, with few exceptions including China, being the original source, and, to a lesser extent, the early involved Italy \cite{LVANDORP}.  Quantitatively, amino acid mutations were found to be significantly more frequent over the entire viral sequence in SARS-CoV-2 genomes tracked in Europe (43.07\%), than in Asia (38.08\%) and in North America (29.64\%) \cite{MRISLAM}.

Here we compare sequences of single filament RNA SARS CoV-2 viruses coming from different countries, using Chern-Simons currents to potentially explain the reason why SARS-CoV-2 variants seem to exhibit a higher incidence during the 2020/2021 pandemic. Finally in Sec. \ref{sect5} we conclude the work discussing results and future perspectives.

\section{The Chern-Simons Theory for DNA Systems} \label{sect2}

In this section we overview the application of Chern-Simons theory to DNA/RNA systems, outlining the main results obtained in \cite{Capozziello:2017zfk}. The first step is to use quaternion fields to define a set of Nitrogen Bases over the DNA or RNA, namely
\begin{equation}
\begin{cases}
\displaystyle A_{DNA} := e^{\frac{\pi}{2} i \beta_n} \;\;\;\;\;\;\;\;\; A_{RNA} := e^{\frac{\pi}{2} j \alpha_n}
\\
\displaystyle T_{DNA}:= i \, e^{-\frac{\pi}{2} i \beta_n}  \;\;\;\;\;\; U_{RNA} := i \, e^{-\frac{\pi}{2} j \alpha_n} 
\\
\displaystyle C_{DNA}:= j \, e^{i \pi \beta_n} \;\;\;\;\;\;\;\; C_{RNA}:= j \, e^{j \pi \alpha_n} 
\\
\displaystyle G_{DNA}:= k \, e^{ 2\pi i \beta_n} \;\;\;\;\;\; G_{RNA}:= k \, e^{ 2\pi j \alpha_n},
\end{cases}
\label{ATCG}
\end{equation}
being  $[h] \in \mathbb{H}$: $[h]= a + b\,i + c\, j + d\,k$ and $a, b, c, d \, \in \mathbb{R}$. The one-form connection $\textbf{A}$ can be thought as a state of the above written nitrogen bases, namely $\textbf{A} \in \left\{A,T/U,C,G\right\}$); consequently the DNA curvature in the configuration space of nitrogen bases is represented by the two-form curvature $F = d\textbf{A}$, which in coordinates representation can be written as:
\begin{equation}
F_{\mu \nu} = \partial_{[\mu} A_{\nu]} + A_{[\mu} A_{\nu]}.
\end{equation}
Therefore, taking into account the $SU(2)$-invariant Chern-Simons three-dimensional action 
\begin{equation}
S^{SU(2)} = \int \text{Tr} \left[\textbf{A} \text{d} \textbf{A} + \frac{2}{3} \textbf{A} \wedge \textbf{A} \wedge \textbf{A} \right],
\end{equation}
it is possible to define the \emph{Chern-Simons current} as the measurable, gauge invariant quantity that can be obtained from the expectation value of the Wilson loop:
\begin{equation}
J = < \left[W(\textbf{A}) \right]> = \frac{\int \mathcal{D} A \, e^{i S} \Pi_n W(A_n)}{\int  \mathcal{D} A \, e^{i S}}.
\end{equation}
Wilson loop is the trace of a path-ordered exponential of the gauge connection and represents the only gauge invariant of the theory:
\begin{equation}
 W(\textbf{A})  = \text{tr} \left[\text{exp} \left\{ \mathcal{P} \oint \textbf{A} \right\} \right] .
\end{equation} 
Wilson Loops can be obtained from the holonomy of the gauge connection around a given loop and are mainly used in gauge lattice theories and quantum chromodynamics \cite{Maldacena:1998im, Lepage:1992xa, Aoki:2006br, Appelquist:1981vg}. They have been formerly introduced to address a nonperturbative formulation of quantum chromodynamics \cite{Wilson:1974sk}, but nowadays play an important role in the formulation of Loop Quantum Gravity, particle physics and String Theory.

The choice of the three-dimensional action is the key point of the method: standard biology suggests that nitrogen bases combine each other in triplets, forming therefore a three-dimensional space of configurations that can be described by means of the Chern-Simons three form. Any point of the space is, thus, labeled by a given triplet. Sixty-four possible combinations arise after combining the nitrogen bases in triplets, and correspond to the combinations occurring in the genetic code. For this reason the space turns out to be discrete and finite.

By means of Eq. \eqref{ATCG}, it is possible to define a discrete superstate of configurations, in which the nitrogen bases represent the dynamical variables, so that the genetic code is labeled by the Chern-Simons currents only. After few calculations, the curvature spectrum of the genetic code can be obtained \cite{Capozziello:2017zfk}, as reported in \textbf{Table 1}.
\newpage
\begin{center}
\textbf{Table 1.} Value of Chern-Simons current for the triplets of the genetic code.
\end{center}
\begin{centering}
\begin{tabular}{l c c c c c c c}\hline\hline
\multicolumn{1}{c}{\textbf{Amino acid}} & \textbf{\, CS Current} & \textbf{\,Amino acid} &\textbf{\,CS Current} & \textbf{\, Amino acid} &\textbf{\, CS Current} & \textbf{\, Amino acid} &\textbf{\, CS Current} \\ \hline
Phe (UUU) & 0.7071 & Ser (UCU)& 0.0534 & Tyr (UAU)& 0.0214 & Cys (UGU)& 0.0122 \\
Phe (UUC) & 0.5000 & Ser (UCC)& 0.0495 & Tyr (UAC)& 0.0205 & Cys (UGC)& 0.0118 \\
Leu (UUA) & 0.3717 & Ser (UCA)& 0.0460 & Sto (UAA)& 0.0197 & Sto (UGA)& 0.0115 \\
Leu (UUG) & 0.2887 & Ser (UCG)& 0.0429 & Sto (UAG)& 0.0189 & Trp (UGG)& 0.0112 \\
Leu (CUU) & 0.2319 & Pro (CCU)& 0.0402 & His (CAU)& 0.0182 & Arg (CGU)& 0.0109 \\
Leu (CUC) & 0.1913 & Pro (CCC)& 0.0377 & His (CAC)& 0.0175 & Arg (CGC)& 0.0106 \\
Leu (CUA) & 0.1612 & Pro (CCA)& 0.0354 & Gin (CAA)& 0.0169 & Arg (CGA)& 0.0103 \\
Leu (CUG) & 0.1382 & Pro (CCG)& 0.0334 & Gin (CAG)& 0.0163 & Arg (CGG)& 0.0010 \\
Ile \,\,\,(AUU)& 0.1201 & Thr (ACU)& 0.0316 & Asn (AAU)& 0.0157 & Ser (AGU)& 0.0098 \\
Ile \,\,\,(AUC)& 0.1057 & Thr (ACC)& 0.0299 & Asn (AAC)& 0.0152 & Ser (AGC)& 0.0096 \\
Ile \,\,\,(AUA)& 0.0939 & Thr (ACA)& 0.0284 & Lys (AAA)& 0.0147 & Arg (AGA)& 0.0093 \\
Met (AUG)& 0.0841 & Thr (ACG)& 0.0270 & Lys (AAG)& 0.0142 & Arg (AGG)& 0.0091 \\
Val \,(GUU)& 0.0759 & Ala (GCU)& 0.0257 & Asp (GAU)& 0.0138 & Gly (GGU)& 0.0089 \\
Val \,(GUC)& 0.0690 & Ala (GCC)& 0.0245 & Asp (GAC)& 0.0134 & Gly (GGC)& 0.0087 \\
Val \,(GUA)& 0.0630 & Ala (GCA)& 0.0234 & Glu (GAA)& 0.0129 & Gly (GGA)& 0.0085 \\
Val \,(GUG)& 0.0579 & Ala (GCG)& 0.0224 & Glu (GAG)& 0.0126 & Gly (GGG)& 0.0083 \\
\hline\end{tabular}
\end{centering}
\vspace*{0.5cm}

The same analysis can be also pursued by considering the amino acids, so that the genetic code is equivalently described by 21 different Chern-Simons currents. The simplest way to construct a curvature spectrum with respect to amino acids, is to take the average values of the Chern-Simons currents which refer to triplets coding for the same amino acid. The Chern-Simons currents of the amino acids are listed in \textbf{Table 2}. 

\begin{center}
\textbf{Table 2.} Value of Chern-Simons current for the amino acids.
\end{center}
\begin{centering}
\begin{tabular}{l c c c c c c c}\hline\hline
\multicolumn{1}{c}{\textbf{Amino acid}} & \textbf{\, CS Current} & \textbf{\,Amino acid} &\textbf{\,CS Current} & \textbf{\,Amino acid} &\textbf{\,CS Current} & \textbf{\,Amino acid} &\textbf{\,CS Current} \\ \hline
Phe (F) & 0.60355 & Ser (S) & 0.0352 & His (H)& 0.01785 & Giu (E)& 0.01275 \\
Leu (L) & 0.2305 & Pro (P) & 0.036675 & Gin (Q)& 0.0166 & Cys (C)& 0.012 \\
Ile \,\,\,(I) & 0.106567 & Thr (T)& 0.029225 & Asn (N)& 0.01545 & Trp \, (W)& 0.0112 \\
Met (M) & 0.0841 & Ala (A)& 0.024 & Lys (K)& 0.01445 & Arg (R) & 0.01005 \\
Val \,\,(V) & 0.06645 & Tyr (Y) & 0.02095 & Asp (D) & 0.0136 & Gly (G)& 0.0086 \\
\hline\end{tabular}
\end{centering}
\vspace*{0.5cm}

Notice that the formalism permits to assign a numerical value to each component of the genetic code, finding a point by point correspondence between triplets and curvature. Such a curvature of the DNA is the key parameter of our approach, as it may provide several predictions about the docking between two different parts of DNA or between DNA and RNA. The genomic curvature can be also used to find out those positions having highest probability to exhibit a mutation. The introduction of the mutation, indeed, leads to a local variation of the curvature, whose value might suggest the clinic importance and the impact of the corresponding disease. Moreover, the curvature spectrum can provide important insights regarding the evolution of the genomic strings: those points with highest curvature are the best candidates to evolve toward a stabler configuration, making the entire sequence more uniform in the configuration space of all the possible triplets.
\section{Application of the Chern-Simons Theory to Biological Systems} \label{sect3}

\subsection{The Chern-Simons Current in Mutated KRAS Human Gene}
The first application of the above described method is focused on the comparison between mutated and standard DNA sequences. In particular, first we consider the KRAS gene\index{KRAS Gene}, whose details are reported in App. \ref{APPA}. It is located in the 12th chromosome, from the base 25,205,246 to 25,250,929 and represents one of the most mutated human genes \cite{hartman, Misale, Lievre}. Then we introduce some known mutations into the original sequence, causing a change in the Chern-Simons current. Being the current linked to the curvature of the DNA, the configuration space made of nitrogen bases changes the point-like curvature wherever a mutation is introduced. 

By means of physical considerations, we theoretically expect the mutation to level out the graph, providing smoother variations of the current with respect to those of the original sequence. By analogy with other physical systems, the curved point is surrounded by a non-equilibrium region, which in turn tends to mutate in order to reach a minimum free energy state.

Moreover, this prescription is in agreement with the general criterion which governs thermodynamic transformations, according to which any spontaneous transformations must minimize the Gibbs free energy. This statement can be simply proved by considering the definition of the Gibbs free energy $\mathcal{G}$, that is
\begin{equation}
    \G = U - T S + pV,
\end{equation}
with $p$ being the pressure, $V$ the volume, $T$ the temperature, $S$ the entropy and $U$ the free energy. Neglecting the contribution of $p$ and setting $T = \text{const.}$ (as standard for biological systems), it turns out that for the system to undergo a spontaneous transformation, the entropy must increase as the free energy decreases. The latter can be thought as the expectation value of the Hamiltonian of the system, which includes potential and kinetic energies. Therefore, requiring the Gibbs free energy to decrease spontaneously is equivalent to require the gravitational potential to decrease spontaneously. This means that, as the system evolves toward a configuration with $\Delta \G < 0$, the potential energy decreases. By applying these considerations to the formalism developed in Sec. \ref{sect2}, a spontaneous transformation must yield an evolution of the system toward flat regions in the configuration space.  

For these reasons, mutations of DNA/RNA sequences occur to render the graph smoother and to bring the general state toward an equilibrium configuration. Reversing the argument, those mutations which make the sequence more peaked than the original one, are supposed to occur less frequently, since they lead to a higher free energy configuration. Therefore, significant variation should not occur in flat regions of the curvature spectrum, which are closer to an equilibrium state. The result of the analysis in KRAS human gene via Chern-Simons current method is reported in \textbf{Fig. 1a}.

\begin{center}
\centering
\includegraphics[width=.99\textwidth]{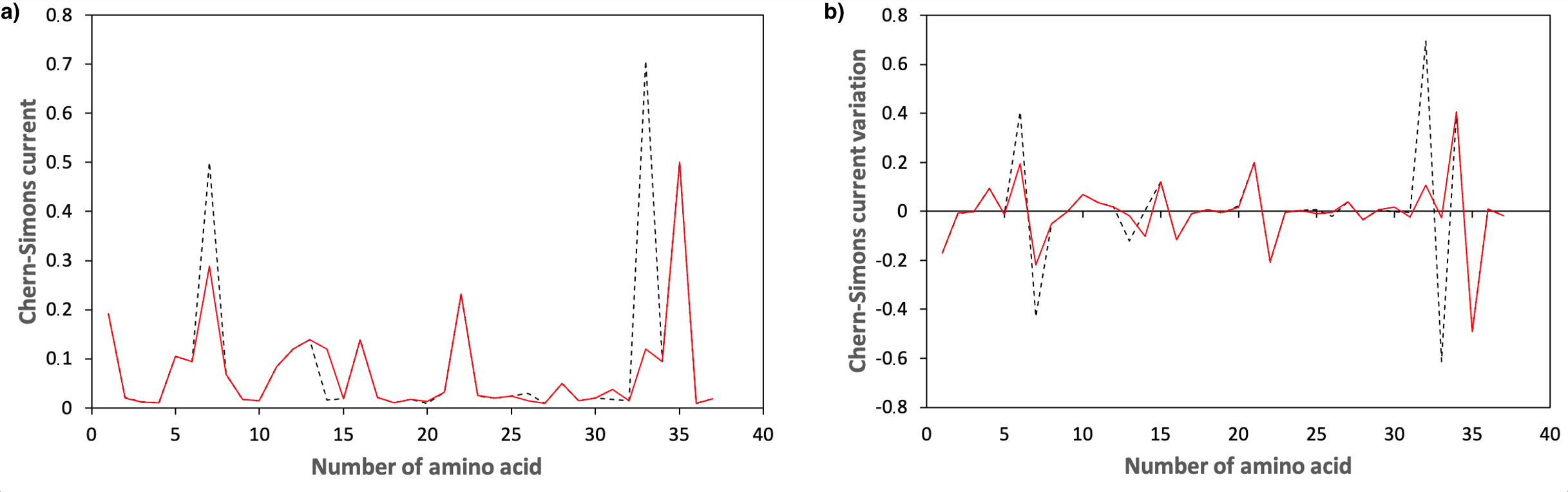}
\\ \textbf{Figure 1}. Chern-Simons current in KRAS gene. Figure 1a shows the comparison between the original sequence (black dashed line) and the mutated one (red solid line), while Figure 1b shows the Chern-Simons current variation, obtained comparing the point-like differences between contiguous points of the original and mutated sequences. The region considered is 25,245,274 - 25,245,384 of the 12th chromosome.
\end{center}
Most significant mutations occur in the regions comprised between the 5th and the 15th amino acid, and between the 30th and the 35th. For this reason, within these intervals, the original sequence differs from the mutated one. This is due to the fact that the presence of the point-like mutation (see \emph{e.g.} position 7, 14 and 33) also influences the curvature of the surrounding regions. Nevertheless, the Chern-Simons currents of the two sequences converge again in correspondence of those points which are not affected by mutations. This shift between original and mutated sequence is more evident in \textbf{Fig. 1}, due to the large amount of mutations introduced in a short sequence made of few amino acids (see \textbf{Table 3}). Further details are reported in App. \ref{APPA}. 

As expected by the free energy minimization argument, mutations occur whereas the curvature is most peaked, providing a smoother general trend, with respect to the original one. Notice, however, that mutations are not directly correlated to peaks, but rather to curvature gradients, namely they are mostly located near those points whose curvature is very much higher (or lower) than their contiguous. By computing the differences between contiguous points, it is possible to associate mutations to peaks, as reported in \textbf{Fig. 1b}.

In the same region of the twelfth chromosome, another set of mutations occurs (\textbf{Fig. 2})
\begin{center}
\centering
\includegraphics[width=.99\textwidth]{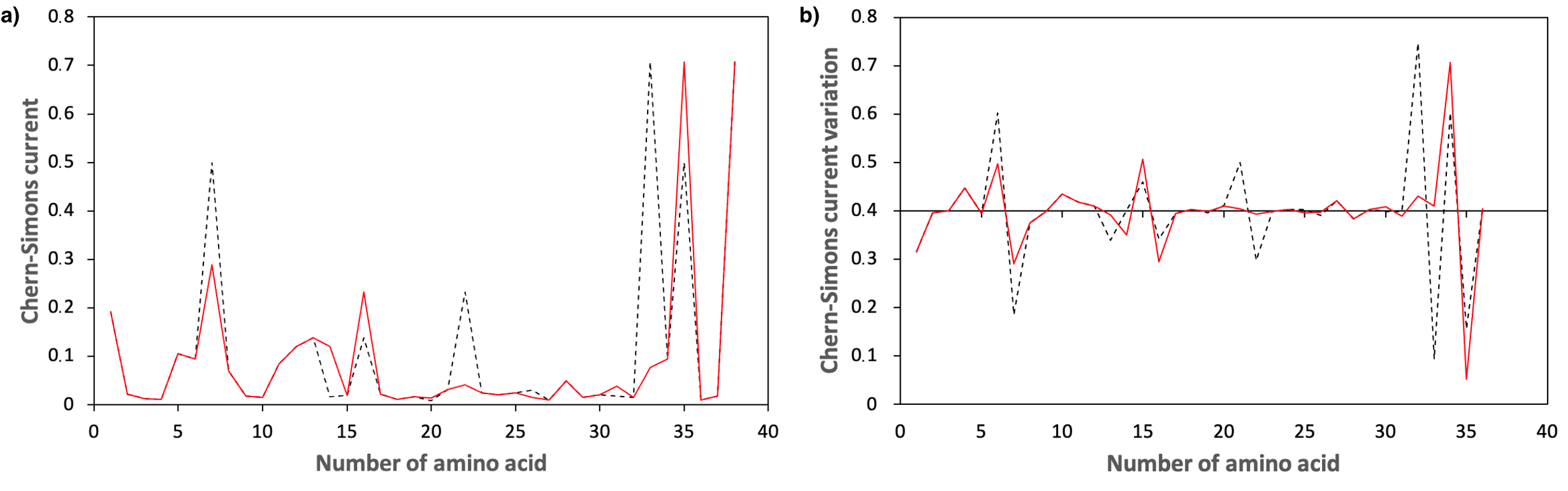}
\\ \textbf{Figure 2.} Chern-Simons current in KRAS gene. Figure 2a shows the comparison between the original sequence (black dashed line) and the mutated one (red solid line), while Figure 2b shows the Chern-Simons current variation, obtained comparing the point-like differences between contiguous points of the original and mutated sequences. The region considered is 25,245,274 - 25,245,384 of the 12th chromosome.
\end{center}

\textbf{Fig. 1} and \textbf{Fig. 2} refer to the same region of KRAS, though different mutations are introduced in the two cases. More precisely, mutations occurring in these selected regions are split in two different sets, in order to facilitate reading and visualizing the curvature spectrum. 

In the second half of the plot, the mutated sequence results shifted with respect to the original one. This can be physically motivated by considering the features of the mutations introduced in position 22 and 26. Specifically, both mutations (see \textbf{Table 4} for details) provide Chern-Simons current values which largely differ from the corresponding original ones. Therefore, though the variation is point-like, the overall trend is highly influenced by the occurrence of such mutations, with the consequence that also the surrounding regions result shifted. However, in position 21 and in position 23 (where no mutations occur) original and mutated sequences have the same current again.

It is worth noticing that, even in this case, a mutation corresponds to each peak, as theoretically inferred. Moreover, the mutated sequence makes the overall trend smoother than the original one, in agreement with theoretical predictions. To confirm this result, two other different regions of human KRAS are analyzed in \textbf{Fig. 3} and \textbf{Fig. 4}, where the original sequences are again compared with the corresponding mutated ones. Mutations are carefully chosen according to the database \href{https://hive.biochemistry.gwu.edu/biomuta/proteinview/P01116}{BioMuta}. Also in this case, further details can be found in App. \ref{APPA}.
\begin{center}
\centering
\includegraphics[width=.99\textwidth]{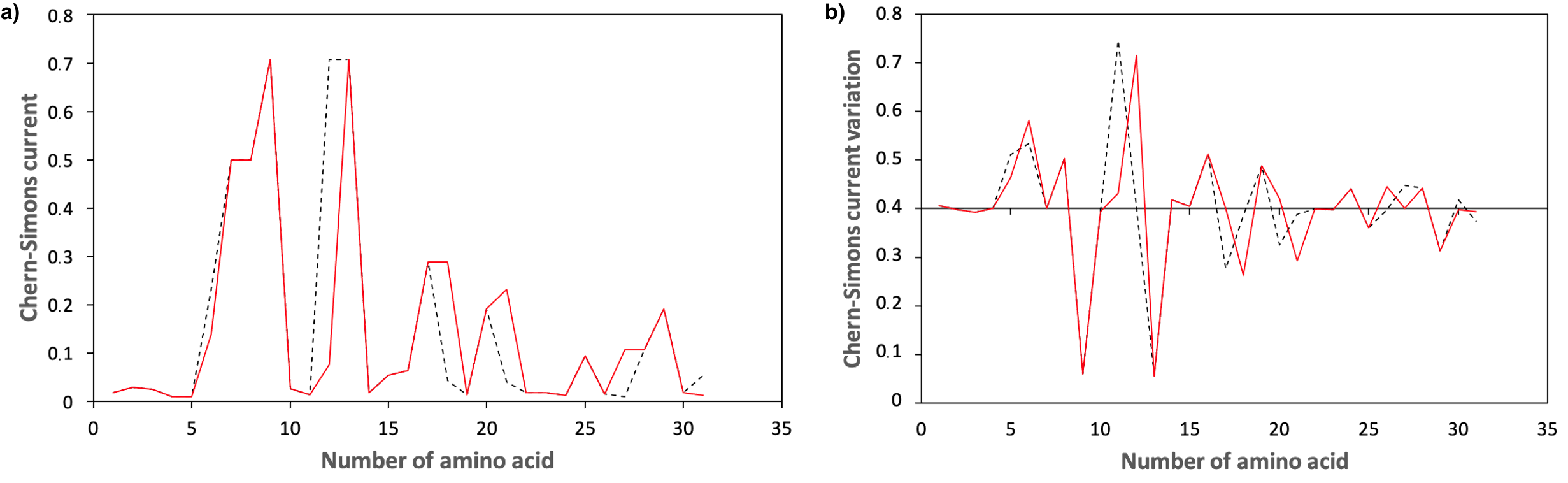}
\\ \textbf{Figure 3.} Chern-Simons current in KRAS gene. Figure 3a shows the comparison between the original sequence (black dashed line) and the mutated one (red solid line), while Figure 3b shows the Chern-Simons current variation, obtained comparing the point-like differences between contiguous points of the original and mutated sequences. The region considered is 25,215,468 - 25,215,560 of the 12th chromosome.
\end{center}

The last region analyzed, corresponding to the region 25,227,263-25,227,379 of the 12th chromosome, yields the graph in \textbf{Fig. 4}.
\begin{center}
\centering
\includegraphics[width=.99\textwidth]{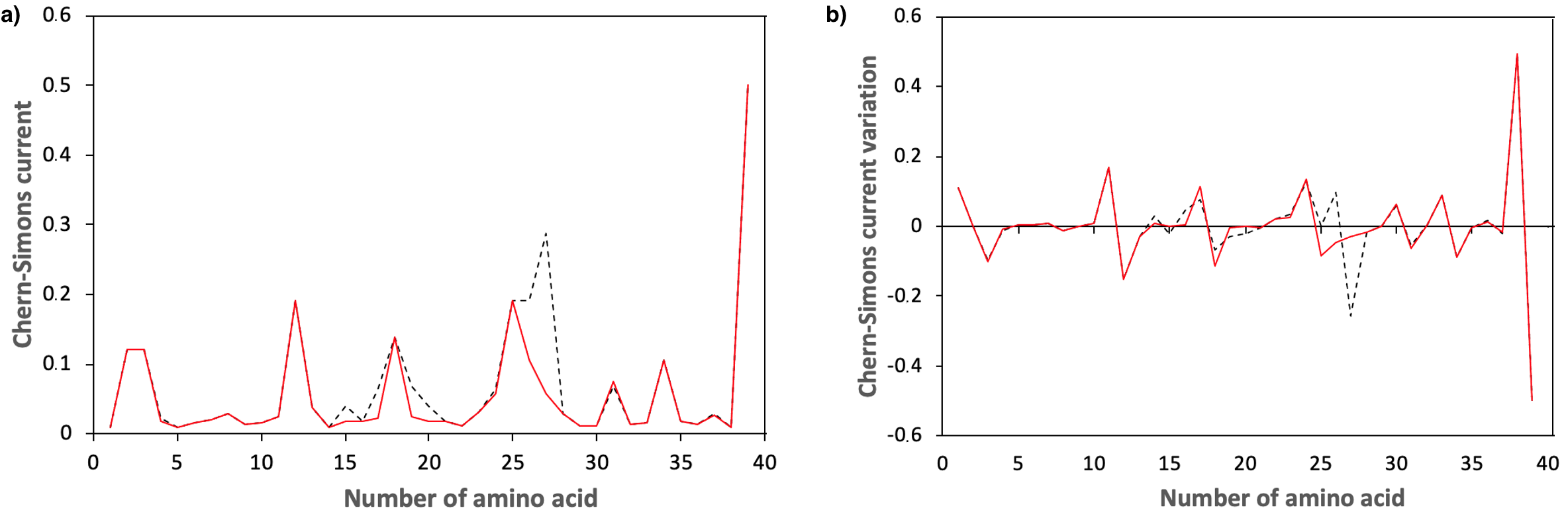}
\\ \textbf{Figure 4.} Chern-Simons current in KRAS gene. Figure 4a shows the comparison between the original sequence (black dashed line) and the mutated one (red solid line), while Figure 4b the Chern-Simons current variation, obtained comparing the point-like differences between contiguous points of the original and mutated sequences. The region considered is 25,227,263-25,227,379 of the 12th chromosome.
\end{center}

Notice that, in both cases, mutations occur where the sequence is peaked, in agreement with theoretical predictions. This is particularly evident in the former case (\textbf{Fig. 3}), where almost all peaks correspond to a mutation (see also \textbf{Table 6}). Moreover, the introduction of the mutations has the effect to avoid abrupt differences in the overall trend of the curvature spectrum. 

On the contrary, notice that few mutations also occur in flat regions. This may be due to other factors that induce mutations, not taken into account by our model at the moment, which is only based on the curvature gradient variation and the free energy minimization.
\subsection{The Chern-Simons Current in Mutated COVID-19 Sequences}
In this subsection we discuss the results provided by the application of the Chern-Simons formalism to different variants of SARS-CoV-2 virus. Let us start by introducing the main features of the latter.

The S glycoprotein is a Class I fusion protein, composed by two subunits (S1,S2) \cite{YYUAN}; the S1 subunit contains the receptor binding domain (RBD), directly binding to the main receptor human angiotensin-converting enzyme 2 (hACE2) and determinant for both host range and cellular tropism \cite{RYAN}; the S2 subunit is directly involved in membrane fusion and virus endocytosis \cite{ACWALLS, XOU}. Receptor binding triggers conformational changes; specifically, host proteases (such as furin) will mediate its functional transition by cleaving the interface between the two subunits (S1, S2). Additionally, the RBDs of SARS-CoV and SARS-CoV-2 are highly similar, despite few key residues, appearing to enhance the transmissibility of the novel CoV \cite{RLU, DWRAPP}. 
The spike glycoprotein is the main inducer for neutralizing antibodies \cite{CADEHAAN}; unwillingly, it shows the highest mutation rate among SARS-CoV-2 proteins \cite{SMLOKMAN, SKUMAR}, and a variable glycosylation can create novel CTL epitopes, possibly altering hACE2 binding and accessibility to proteases and neutralizing antibodies \cite{ACWALLS, JLAN}. 

The purpose here is to find a correlation in terms of Chern-Simons current among the mutations of the sequences, a correlation that could possibly give insights aiming at localizing and predicting mutation sites in the new variants of the virus. We analyze eleven strings, which underwent mutations with respect to the original sequence of SARS-CoV-2, firstly detected in Wuhan at the end of 2019. They all correspond to the same RNA region and was selected in accordance with \textbf{Fig. 5}. In particular, we compare the difference of Chern-Simons currents, considering variants from Asia, Europe, Oceania and North America. Specifically, sequence 19A is the first one which arose in Wuhan and have been spreading during the initial 2020 outbreak; 19B is the first detected variant in China; 20A dominated mostly in Europe from march 2020, to subsequently spreading out globally; 20B and 20C are variants of 20A which mainly spread in the early 2020; finally, 20D, 20E, 20F, 20G, 20H, 20I occurred on summer 2020 as variants of 20B, 20C and 20A. Among them, 20I and 20H are English and south-African variants. To be more precise, we used the tool \href{https://clades.nextstrain.org}{Nextclade}, yielding the graph of \textbf{Fig. 5}. This figure shows the aforementioned evolution of the sequences (\href{https://github.com/nextstrain/ncov/blob/master/defaults/clades.tsv}{https://github.com/nextstrain/ncov/blob/master/defaults/clades.tsv}).

\begin{center}
\centering
\includegraphics[width=.75\textwidth]{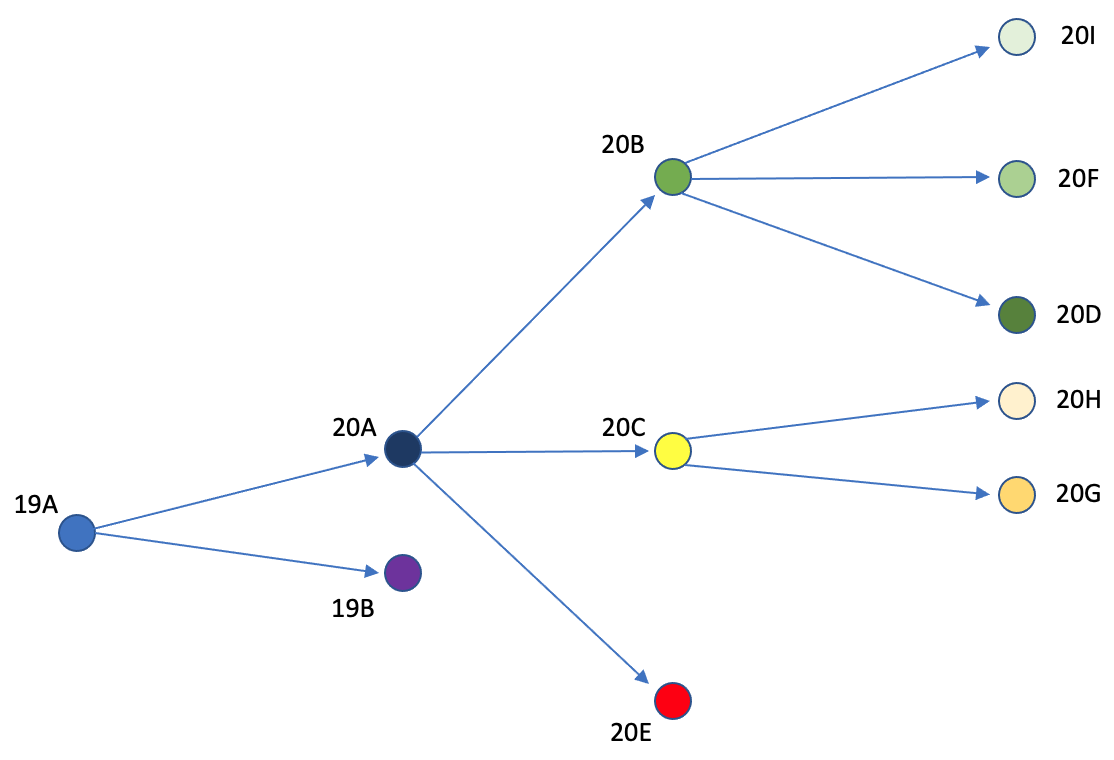}
\\ \textbf{Figure 5.} Evolution of the first-detected Wuhan sequence (19A) to other variants which spread out during the 2020 pandemic.
\end{center}

Mutations of the triplets which caused the occurrence of variants are reported in App. \ref{APPB}. In our analysis, because of the large amount of nitrogen bases, we only compute the difference of Chern-Simons currents between the original sequence and the mutated one. Specifically, we consider the slope of the current for each mutation, namely the number
\begin{equation}
\text{Slope} = \frac{\text{Mutated Seq.}-\text{Original Seq.}}{\text{Original Seq.}}.
\label{slope12}
\end{equation}
Specifically, high values of the slope represent a large discrepancy between the original sequence and the mutated one in the curvature spectrum, while lower values account for small differences. We perform the one-to-one comparison between contiguous sequences (showed in \textbf{Fig. 5}), with the aim to find out a correlation between slopes and mutations. 
Each variant is compared with the corresponding predecessor, so that no comparison is carried on between sequences which are not directly evolving from one another, according to \textbf{Fig. 5}. For example, sequence 19A is not compared with 20I, as well as 20D is not compared with 20H. 

The analysis shows that mutations occur with highest probability where the slope (as defined in Eq. \eqref{slope12}) of Chern-Simons current assumes extreme values, namely when its modulus is extremely high or extremely low\footnote{As reported at the beginning of App. \ref{APPB}, we define current variations as "low" if they are included in the range [-11\%,11\%], and as "high" if they are > 100\% or < - 100\%. Also notice that there is no upper limit to the modulus of the current variation, since it represents the percentage of current increase with respect to surrounding points}.

This means that even those mutations which do not cause significant current variations can support variants. In particular, the one-to-one comparison between the original and the corresponding mutated sequences shows that approx 70\% of mutations corresponds to extreme values of current. Such percentage increases up to 80\% if we consider only those mutations which will effectively spread out (denoted in italic bold and highlighted in light yellow), as showed in App. \ref{APPB}, \textbf{Figs. 7-17}. Consequently, this statistic can be used to point out which occurred mutation of the sequence can be more likely to evolve in a real, spread out variant of the virus. To be more precise, once we know the position of a given mutation, Chern--Simons currents can suggest which type of triplets will arise from such mutation. In particular, as provided by the analysis, the mutated sequences should exhibit mutations whose related Chern-Simons currents provide extremely high or extremely low percentage variations, with respect to the original ones. Therefore, we do not expect the sequence to evolve such that mutations cause intermediate values of current variations; rather, if the position of the mutation is known, we expect the triplet to mutate towards those possible configurations whose Chern-Simons current is either very close or very far from the initial one (in terms of percentage). This means that from a given triplet we can select a set of possible mutations, namely those which cause either high or low current variations.

The above results constitute a part of the analysis of SARS-Cov-2 virus, which mainly relies on the evolution of given sequences towards mutated configurations. As mentioned above, this first part turns out to be useful to restrict all possible mutations within a given range, but can provide suitable information only if the position of the mutation is known \emph{a priori}. From this point of view, no information regarding the mutation position can be provided. Now, in the next part, we use Chern-Simons formalism to select regions where mutations are most likely to occur.

With the aim to link the currents with the probability to exhibit mutations, we analyze only those sequences which generate variants, \emph{i.e.} 19A, 20A, 20B and 20C. Specifically, as we can infer from \textbf{Fig. 5}, 19A generates 19B and 20A; 20A generates 20B, 20C and 20E; 20C generates 20H and 20C. Similarly to the previous analysis of KRAS human gene, we aim to relate the curvature spectrum with the likelihood to find out mutations. To this purpose, we calculated the Chern--Simons currents of 19A, 20A, 20B and 20C sequences and computed the current variations in those points affected by known mutations. Specifically, let $n$ be the position of a given mutation along the sequence and $j_n$ the corresponding Chern--Simons current. The normalized current variations are computed in accordance to the formulas:
\begin{equation}
\text{Variation (\%)}_{1} = \frac{j_{n+1} - j_n}{j_n}
\label{var01}
\end{equation}
and 
\begin{equation}
\text{Variation (\%)}_{2} = \frac{j_{n} - j_{n-1}}{j_{n-1}}.
\label{var02}
\end{equation}
This means that we are investigating the current variations where the mutations occur, with respect to the previous and the subsequent points, respectively. The comparison between these values, calculated for the triplets affected by mutations and the surrounding points, can be used to relate the current variation with RNA mutations. 

This prescription is suggested by the analysis performed on human KRAS regions, where it turns out that points far from the equilibrium state in the curvature spectrum are the best candidates to provide mutations. Here, given the large amount of amino acids, the curvature spectrum cannot be computed entirely. For this reason, we only focused on noticeable mutations, namely preferred points which exhibit known triplet variations. 

The analysis again shows that mutations mostly occur where the current variation, as calculated in Eqs. \eqref{var01} and \eqref{var01}, is high-valued. More precisely, in a set of 125 total mutations, 59\% of them (74/125, see \textbf{Tables 7-10}) are located in points where the curvature undergoes abrupt variations. This percentage increases up to 69\%, if only noticeable mutations which had more impact in the development of the corresponding variants are considered. Indeed, among 25 mutations with the greatest impact in generating the variants, 17 exhibit high percentage variations of current with respect to surroundings points. These results are reported in App. \ref{APPB}, \textbf{Tables 7-10}.
 
This result can be explained based on the achievements of the previous section, where non-equilibrium points turned out to be best candidates to provide nitrogen bases mutations. More precisely, large values of the current variations account for peaked regions, which tend to evolve to a lower curvature, that is a lower current. Reversing the argument, large variations of current are exhibited by points which are far from the minimum of energy, which is supposed to occur where the trend is constant.

In this framework, the application of Chern-Simons theory to DNA/RNA systems such as SARS-CoV-2 or KRAS, can give important information about the positions where the mutation is more likely to manifest. The consequent biological impact naturally follows, since this prediction can be used to prevent the occurrence of variants or to know in advance the probability for the sequence to evolve towards another configuration.

Taking into account these results, let us evaluate the spike region of SARS-CoV-2 virus only, with the aim to analyze the tertiary structure. In particular, we rely on the interaction points reported Ref. \cite{Watanabe}, according to which the amino acids of the spike protein are interact as reported in \textbf{Fig. 6} \footnote{Numbers refer to the positions on the spike protein only}.

\begin{center}
\centering
\includegraphics[width=.80\textwidth]{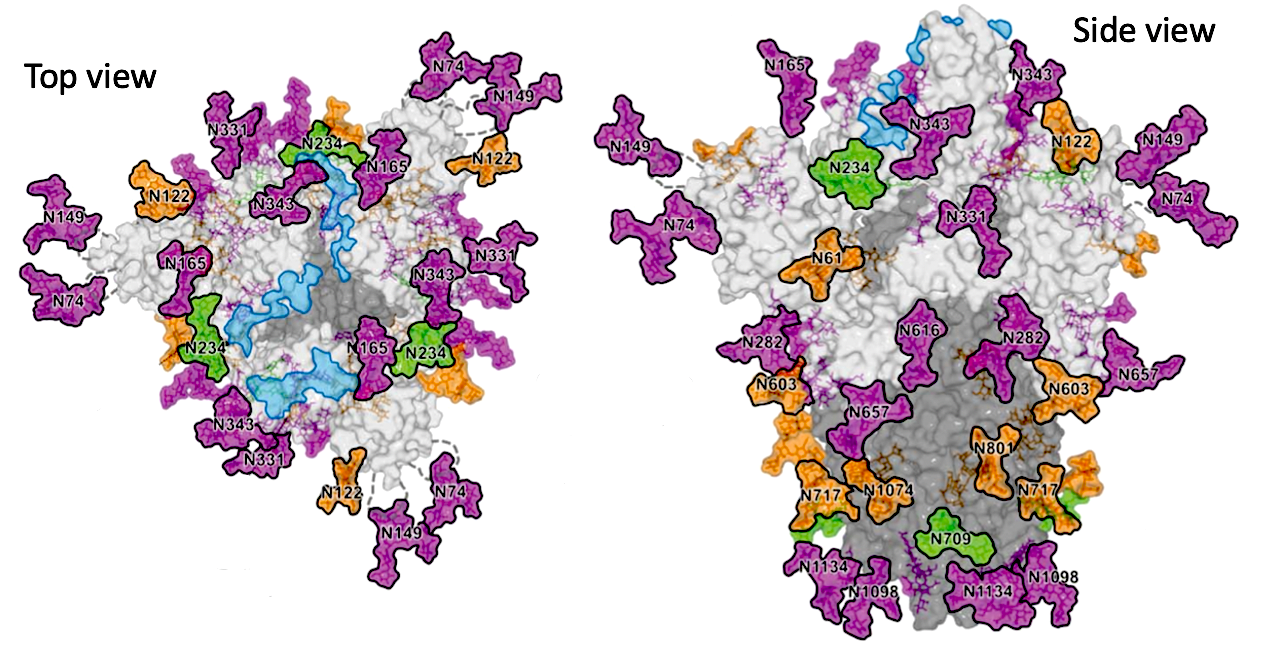}
\\ \textbf{Figure 6.} Tertiary structure of the spike protein of SARS-CoV-2 virus (as taken from \cite{Watanabe}, Fig. 3 therein). Green, orange and pink colors refer to the oligomannose content. Specifically, glycan sites labeled in green contain 80-100\% of oligommannose, those labeled in orange 30-79\% and those labeled in pink 0-29\%. Light blue denotes ACE2 binding sites.
\end{center}

In light of the results provided by Ref. \cite{Watanabe}, we analyzed 11 contact points, namely 22 corresponding amino acids. The features of these latter, such as position, current or percentage variation with respect to the surrounding triplets are reported in \textbf{Table 11}. 

We considered 22 sites and calculated the Chern-Simons current variation of each amino acid with respect to the surrounding points in the linear structure. Beside the first amino acid (position 19), none of them is affected by known mutations. It is interesting to observe that the amount of large variations in those sites which are not affected by mutations is 7/21, namely 33\%. Note that such a percentage is quite lower than the previously discussed one, which is of the order of 72\%. This confirms that Chern-Simons current variations is high-valued whereas mutations develop. Moreover, these seven sites which undergo large percentage variations are oligomannose-type, as pointed out in Ref. \cite{Watanabe}. This, in principle, could be the reason of such large values. For instance, the high value of current variation in position 234 might be due to the proximity of the site with ACE-2, or to the high percentage of glycosylation occurring in such amino acid. 

Moreover, it turns out that the docking points have same or similar values of current, which means low percentage variation. This is expected from a physical point of view, since those points with same curvature tend to interact in order to reach a stabler configuration. Also here, the analogy with gravitational interaction is simply understood. 

\section{Conclusions and Perspectives} \label{sect5}
In this work we apply the schematization method of the nucleic acids representation, based on the Chern-Simons theory as developed in \cite{Capozziello:2017zfk, Capozziello:2018lnd}. Our main purpose is to analyze DNA sequences, such as those contained in the KRAS human gene, and some RNA noticeable sequences such as those of the most known SARS-CoV variants. In particular, we compare known windows of the reference sequences with the corresponding noticeable mutations, reported in well-known and reputed genetic databases. To develop the formalism, the nitrogen bases are recast as quaternion fields, combined in triplets as dictated by biology golden rules. These triplets form a three-dimensional space of configuration that can be described through the Chern-Simons three form. The expectation value of the only observable of the theory, the Wilson Loop, provides the so called Chern-Simons current. The latter gives point-like information of the curvature of the genetic code, and can be used to compute the curvature spectrum of a given genetic string. If some triplet of the initial sequence changes due (for example) to the replacement of a nitrogen base, the point-like curvature changes accordingly. Therefore, the introduction of some mutations yields a variation in the Chern-Simons current. The difference between the original and the mutated sequence can be used to infer where DNA-DNA (or DNA-RNA) interactions take place, or to predict the evolution probability toward a given configuration. 

On the one hand, the latter application of our method can shed light on the possibility to develop proper vaccination strategies against, for instance, SARS-CoV-2 virus; on the other hand it can potentially be used to monitor pharmacological therapies and to quantify the risk of developing DNA/RNA mutations between remission and relapsing phases.
    
The result of the analysis of four different regions of KRAS human gene, an important gene acting as on/off switch in cell signaling and controlling cell proliferation, shows that common features are shared in all analyzed cases. Specifically, in almost all cases, a curvature peak of the regions corresponds to a known mutation, which often yields a new smoother curvature spectrum with respect to the reference. This can be theoretically motivated by physical considerations: the most peaked regions represent non-equilibrium points, which tend to evolve toward stabler configuration of minimum free energy. 

Consequently, it follows that the variations in the curvature spectrum, leading to genetic mutations, likely take place in those regions with higher curvature. This means that, as an effect of the mutations, the overall trend of the curvature spectrum of the sequence tend to become smoother and smoother with no avoid abrupt variations, making nearby points to have similar values of current. As mentioned above, this happens for most of the analyzed cases; however, DNA and RNA evolution can certainly also depend on many other factors that cannot be taken into account by this method. The application of Chern–Simons theory to DNA systems, indeed, only relies on the intrinsic curvature calculation assumed by biological systems in the configuration space made of nitrogen bases. A free energy minimum principle, then, leads to the evolution of the configurations and may suggest likely positions for possible mutations.

We utilize our method also to analyze RNA sequences: in this case we pick the COVID-19 virus, a striking example of the present time, and apply the same prescription to more than 20Kbases of the COVID-19 virus, coming from different countries. Due to the intrinsic attitude of RNA viruses to change their sequence with replication, mutations of various types can occur such as recombination and reassortment, rendering more complex the related genomic analyses. 

Rather than analyzing the entire RNA sequence of the virus, which is very long, we prefer to focus on the regions that are reported to exhibit the most significant mutations, such as the region coding for the SARS-CoV-2 spike protein. Interestingly, the analysis shows that most of mutations occur where the slope of the Chern--Simons current takes extremely high values, which accounts for peaked regions in the curvature spectrum. This result can be explained again considering the principle of minimum free energy, according to which amino acids in correspondence of peaks of the Chern-Simons current value are intrinsically unstable and therefore tend to evolve towards a stabler configuration. Furthermore, we note that few mutations are also exhibited in correspondence of low current values. This may happen because some regions with low current values, namely having a small curvature and being rather flat, often are the border with areas with steep gradients of the current value denoting high curvature. Then, in some cases, even regions with very small curvature may be affected by a close instability, due to the presence of a current gradient nearby. By comparing low current variations listed in \textbf{Figs. 7-16} with \textbf{Tables 7-10}, it turns out that 47\% of points which exhibit low current variations (between mutated and original sequences), are unstable due to the presence of a current gradient nearby.

Notice that, in our analysis, we only considered 2020 sequences of SARS-CoV-2 virus. This is mostly due to the fact that the best part of variants spread out during 2020, thus a comparison like the one reported in \textbf{Fig. 5} turns out to be more interesting.

As a final remark, the importance of the applications here discussed is twofold. On the one hand, it tests the capability of a topological theory in schematizing DNA and RNA configurations to correctly represent their interactions and mutations. On the other hand, it suggests a general criterion to predict the location in genetic sequences where it could be most likely a mutation to take place. This novel method, based on analogue gravity, can be helpful in addressing biological issues, especially when combined with standard bioinformatic approaches. For instance, the probable evolution of a given string, provided by the Chern-Simons formalism, can be approached to mathematical and statistical techniques to increase the likelihood to localize the mutations. In this sense, the approach is deterministic and based on the dynamics of structures, rather than on their mere description. In future works we plan to provide further confirmation of the validity of our approach, by extending the analyses to other genetic sequences both for DNA- and RNA-based systems. We also aim to study the interactions between macro molecules, in order to check whether their point-like curvature values can provide information regarding the docking probability, or predict the points where interactions occur.
\newpage
\section*{Acknowledgements}

C.A., F.B. and S.C. acknowledge the support of  {\it Istituto Nazionale di Fisica Nucleare} (INFN) ({\it iniziative specifiche} MOONLIGHT2 and  GINGER). C.A. acknowledges support from the Italian Ministry for Research under the Project PRIN - Predicting and controlling the fate of bio-molecules driven by extreme-ultraviolet radiation - Prot. Nr.20173B72NB. C.A. and F.B. aknowledge the project PON (Programma Operativo Nazionale Ricerca e Innovazione) 2014-2020 (CCI 2014IT16M2OP005), "Tecnologie innovative per lo studio di interazioni tra acidi nucleici e proteine: metodi sperimentali e modelli"; project code DOT1318991. G.F. acknowledges Giorgio Giurato for the useful discussions and suggestions regarding bioinformatic data. 
L.A. and R.B. aknowledge financial support by the "Associazione Italiana per la Ricerca sul Cancro" (AIRC IG17217 to L.A.), the Italian Ministry for University and Research (PRIN2015-20152TE5PK, to L.A.), the project “Epigenetic Hallmarks of Multiple Sclerosis" (acronym Epi-MS) (id:415, Merit Ranking Area ERC LS) in VALERE 2019 Program (to R.B.) and Programma V:ALERE 2020 - Progetto competitivo “CIRCE" in risposta al bando D.R. n. 138 del 17/02/2020 (to R.B.); Blueprint 282510 (to L.A.); EPICHEMBIO CM1406 (to L.A.); Campania Regional Government Technology Platform Lotta alle Patologie Oncologiche: iCURE (to LA); Campania Regional Government FASE2: IDEAL (to L.A.); MIUR, Proof of Concept POC01\_00043 (to L.A.); POR Campania FSE 2014-2020
ASSE III (to L.A.).

\section*{Author contributions statement}

Conceptualization, F.B., C.A., L.A. and S.C.; Formal analysis, F.B., C.A. and M.D.S.; Methodology, L.A., R.B., M.D.S. and G.F.; Supervision, C.A, L.A., S.C.; Writing-original draft, F.B., G.F. and R.B.; All authors have read and agreed to the published version of the manuscript. All authors reviewed the manuscript. \textbf{}

\newpage
\appendix 
%\lipsum[1-100]
\section{Sequences used and Corresponding Mutations in KRAS}
\label{APPA}

\hspace*{0.36cm} \textbf{KRAS HUMAN}

\textbf{SOURCE FOR THE SEQUENCES}: \href{http://genome-euro.ucsc.edu/cgi-bin/hgTracks?position=KRAS&pix=840&Submit=Submit&db=hg19}{Genome Browser}

\textbf{SOURCE FOR THE MUTATIONS}: \href{https://hive.biochemistry.gwu.edu/biomuta/proteinview/P01116}{BioMuta}
\vspace*{0.5cm}

\textbf{ORIGINAL SEQUENCE 1}: \, Chr12: 25,245,274 - 25,245,384 
\vspace*{0.5cm}

CUCUAUUGUUGGAUCAUAUUCGUCCACAAAAUGAUUCUGAAUUAGCUGU\\AUCGUCAAGGCACUCUUGCCUACGCCACCAGCUCCAACUACCACAAGUUU\\AUAUUCAGUCAU
\vspace*{0.5cm}

\textbf{First set of mutations (\textbf{Fig. 1})}
\vspace*{0.5cm}

\begin{center}
\centering
\textbf{Table 3.} Comparison between original and mutated sequences in KRAS. Chr12: 25,245,274 - 25,245,384
\includegraphics[width=.95\textwidth]{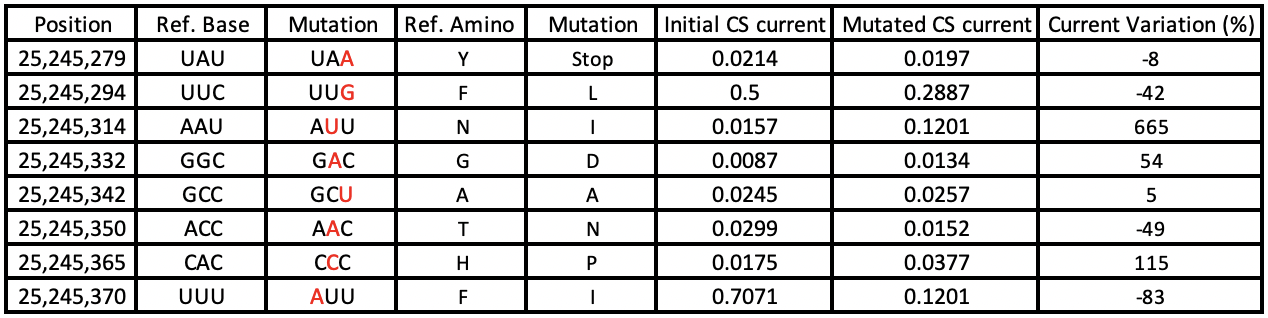}
\end{center}

CUCUA\textcolor{red}{A}UGUUGGAUCAUAUU\textcolor{red}{G}GUCCACAAAAUGAUUCUGA\textcolor{red}{U}UUAGCUGU\\AUCGUCAAG\textcolor{red}{A}CACUCUUGC\textcolor{red}{U}UACGCCA\textcolor{red}{A}CAGCUCCAACUACC\textcolor{red}{C}CAAG\textcolor{red}{A}UU\\AUAUUCAGUCAU 
\vspace*{0.5cm}

\textbf{Second set of mutations (\textbf{Fig. 2})}
\vspace*{0.5cm}

\begin{center}
\centering
\textbf{Table 4.} Comparison between original and mutated sequences in KRAS. Chr12: 25,245,274 - 25,245,384
\includegraphics[width=.95\textwidth]{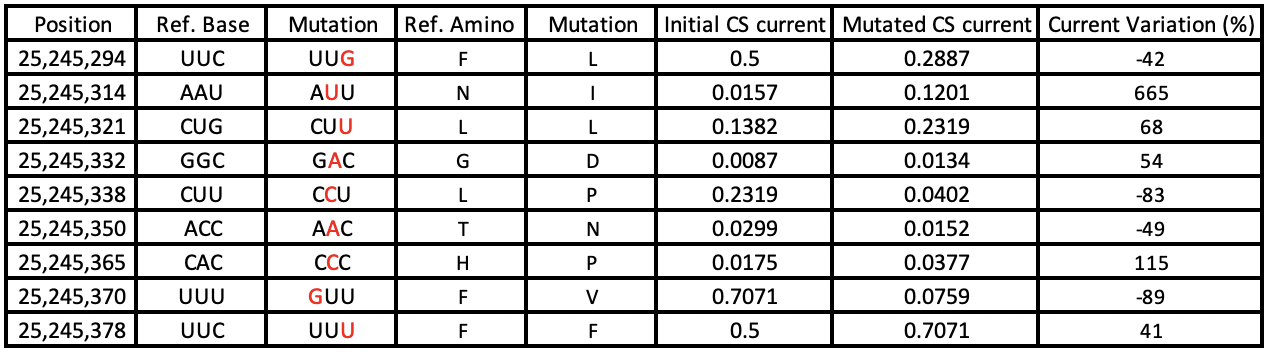}
\end{center}

CUCUAUUGUUGGAUCAUAUU\textcolor{red}{G}GUCCACAAAAUGAUUCUGA\textcolor{red}{U}UUAGCU\textcolor{red}{U}UA\\UCGUCAAG\textcolor{red}{A}CACUC\textcolor{red}{C}UGCCUACGCCA\textcolor{red}{A}CAGCUCCAACUACC\textcolor{red}{C}CAAG\textcolor{red}{G}U\\UAUAUU\textcolor{red}{U}AGUCAU
\vspace*{0.5cm}

\textbf{ORIGINAL SEQUENCE 2}: \,Chr12: 25,215,468 - 25,215,560
\vspace*{0.5cm}

CACACAGCCAGGAGUCUUUUCUUCUUUGCUGAUUUUUUUCAAUCUGUAUU\\GUCGGAUCUCCCUCACCAAU
GUAUAAAAAGCAUCCUCCACUCU
\vspace*{0.5cm}

\textbf{Third set of mutations (\textbf{Fig. 3})}
\vspace*{0.5cm}

\begin{center}
\centering
\textbf{Table 5.} Comparison between original and mutated sequences in KRAS. Chr12: 25,215,468 - 25,215,560
\includegraphics[width=.95\textwidth]{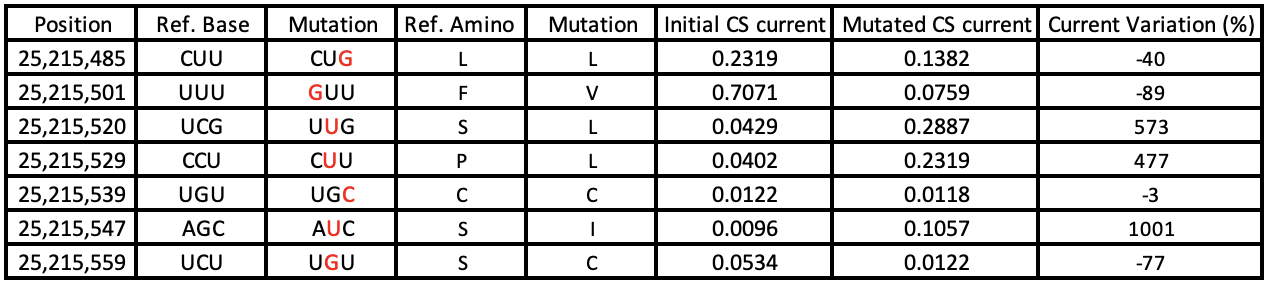}
\end{center}

CACACAGCCAGGAGUCU\textcolor{red}{G}UUCUUCUUUGCUGAU\textcolor{red}{G}UUUUUCAAUCUGUAUU\\GU\textcolor{red}{U}GGAUCUCC\textcolor{red}{U}UCACCAAUG\textcolor{red}{C}AUAAAAA\textcolor{red}{U}CAUCCUCCACU\textcolor{red}{G}U
\vspace*{0.5cm}

\textbf{ORIGINAL SEQUENCE 3}: \,Chr12: 25,227,263 - 25,227,379
\vspace*{0.5cm}

AGUAUUAUUUAUGGCAAAUACACAAAGAAAGCCCUCCCCAGUCCUCAUGUA\\CUGGUCCCUCAUUGCACUGUACUCCUCUUGACCUGCUGUGUCGAGAAUAUC\\CAAGAGACAGGUUUC
\vspace*{0.5cm}

\textbf{Fourth set of mutations (\textbf{Fig. 4})}
\vspace*{0.5cm}

\begin{center}
\centering
\textbf{Table 6.} Comparison between original and mutated sequences in KRAS. Chr12: 25,227,263 - 25,227,379
\includegraphics[width=.95\textwidth]{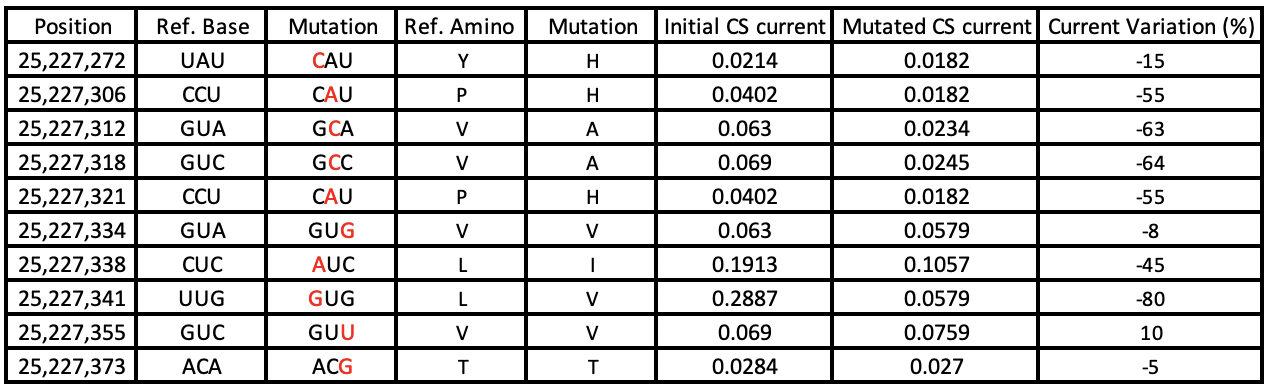}
\end{center}

AGUAUUAUU\textcolor{red}{C}AUGGCAAAUACACAAAGAAAGCCCUCCCCAGUC\textcolor{red}{A}UCAUG\textcolor{red}{C}A\\CUGG\textcolor{red}{C}CC\textcolor{red}{A}UCAUUGCACUGU\textcolor{red}{G}CUC\textcolor{red}{A}UC\textcolor{red}{G}UGACCUGCUGUGU\textcolor{red}{U}GAGAAUAUC\\CAAGAGAC\textcolor{red}{G}GGUUUC
\newpage
\section{Mutations SARS-CoV-2}
\label{APPB}
\subsection*{Comparison Between Original Sequences and Mutated Ones}
The pie graphs of \textbf{Figs. 7-17} show the percentage of large and small values of current variations; large variations ($ >100\% \lor < -100\%$) are labeled by light blue squares, small variations ([-11\%;11\%]) by solid red, other intermediate values by grey lines.
\begin{center}
\centering
\includegraphics[width=.99\textwidth]{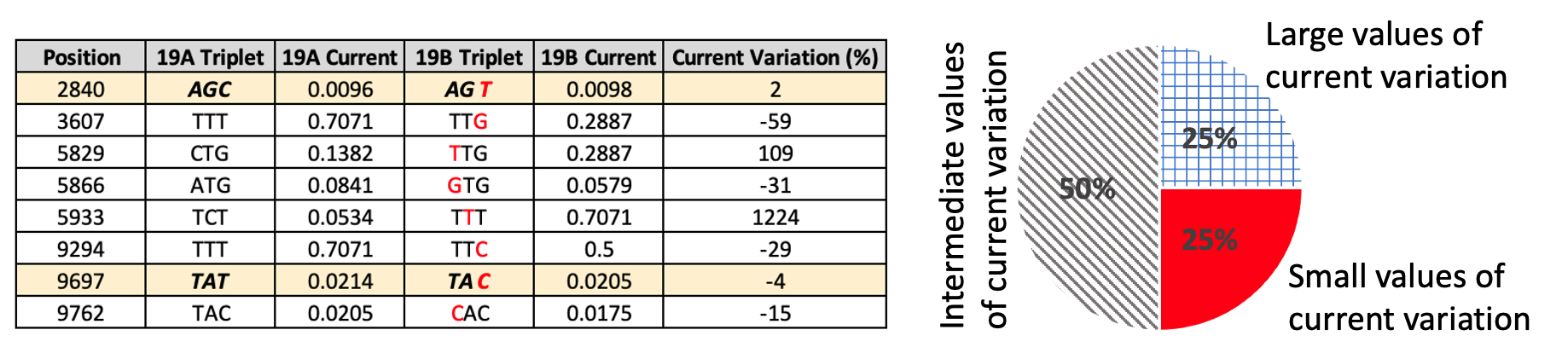}
\\ \textbf{Figure 7.} Comparison between 19A and 19B sequences, with related Chern-Simons current and percentage variation. 
\end{center}

\begin{center}
\centering
\includegraphics[width=.97\textwidth]{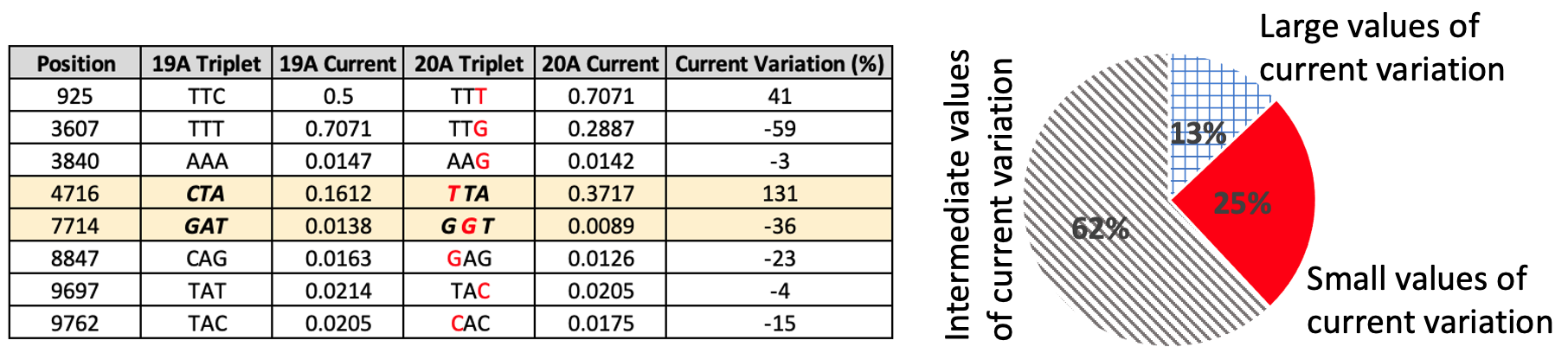}
\\ \textbf{Figure 8.} Comparison between 19A and 20A sequences, with related Chern-Simons current and percentage variation.
\end{center}

\begin{center}
\centering
\includegraphics[width=.97\textwidth]{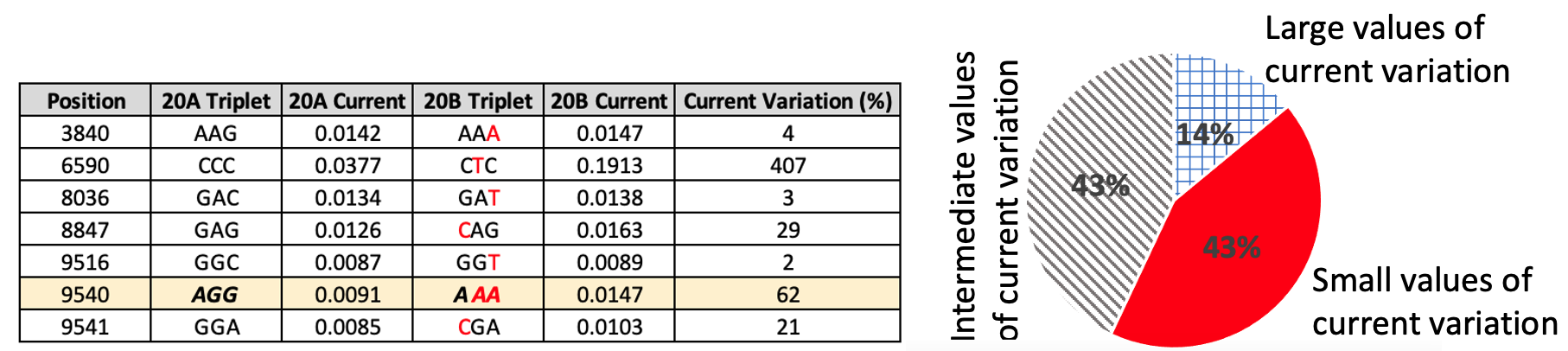}
\\ \textbf{Figure 9.} Comparison between 20A and 20B sequences, with related Chern-Simons current and percentage variation.
\end{center}

\begin{center}
\centering
\includegraphics[width=.98\textwidth]{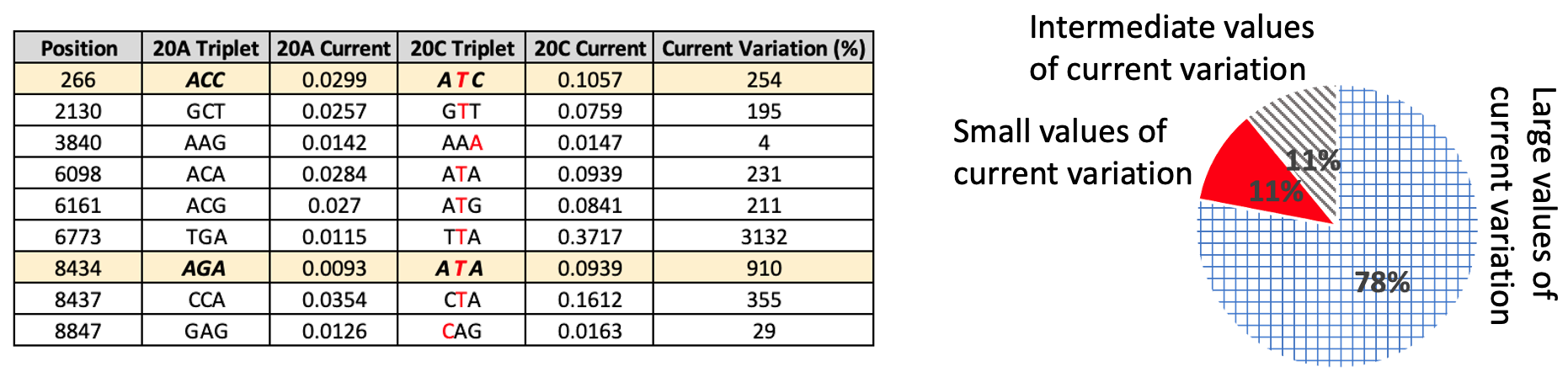}
\\ \textbf{\textbf{Figure 10.}} Comparison between 20A and 20C sequences, with related Chern-Simons current and percentage variation.
\end{center}

\begin{center}
\centering
\includegraphics[width=.97\textwidth]{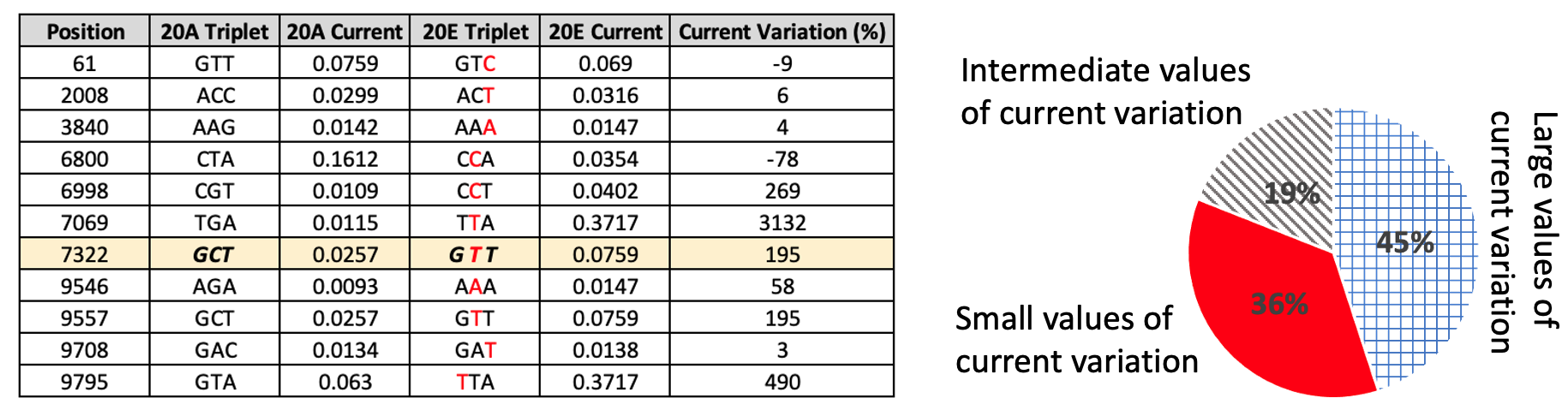}
\\ \textbf{Figure 11.} Comparison between 20A and 20E sequences, with related Chern-Simons current and percentage variation.
\end{center}

\begin{center}
\centering
\includegraphics[width=.97\textwidth]{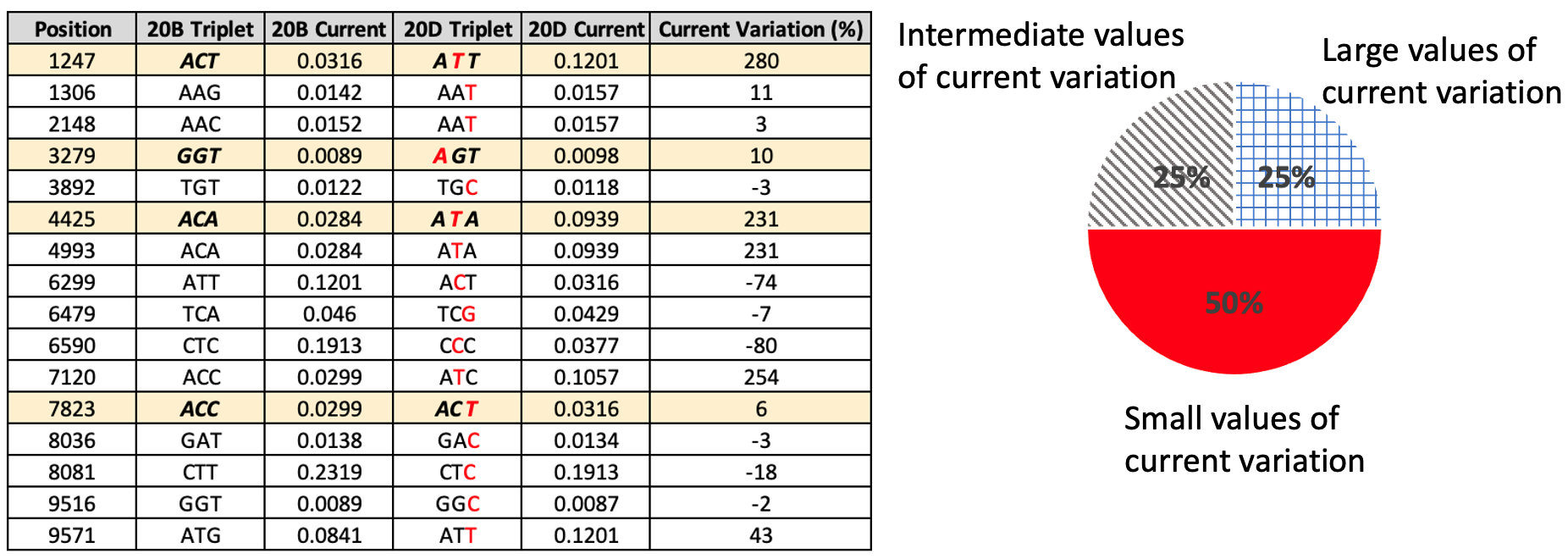}
\\ \textbf{Figure 12.} Comparison between 20B and 20D sequences, with related Chern-Simons current and percentage variation.
\end{center}

\begin{center}
\centering
\includegraphics[width=.99\textwidth]{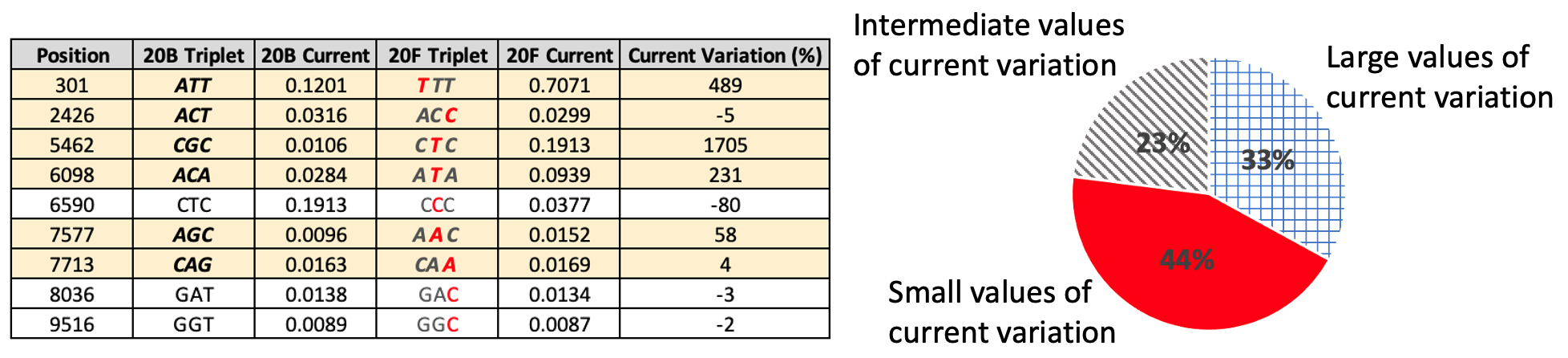}
\\ \textbf{Figure 13.} Comparison between 20B and 20F sequences, with related Chern-Simons current and percentage variation.
\end{center}

\newpage
\begin{center}
\centering
\includegraphics[width=.97\textwidth]{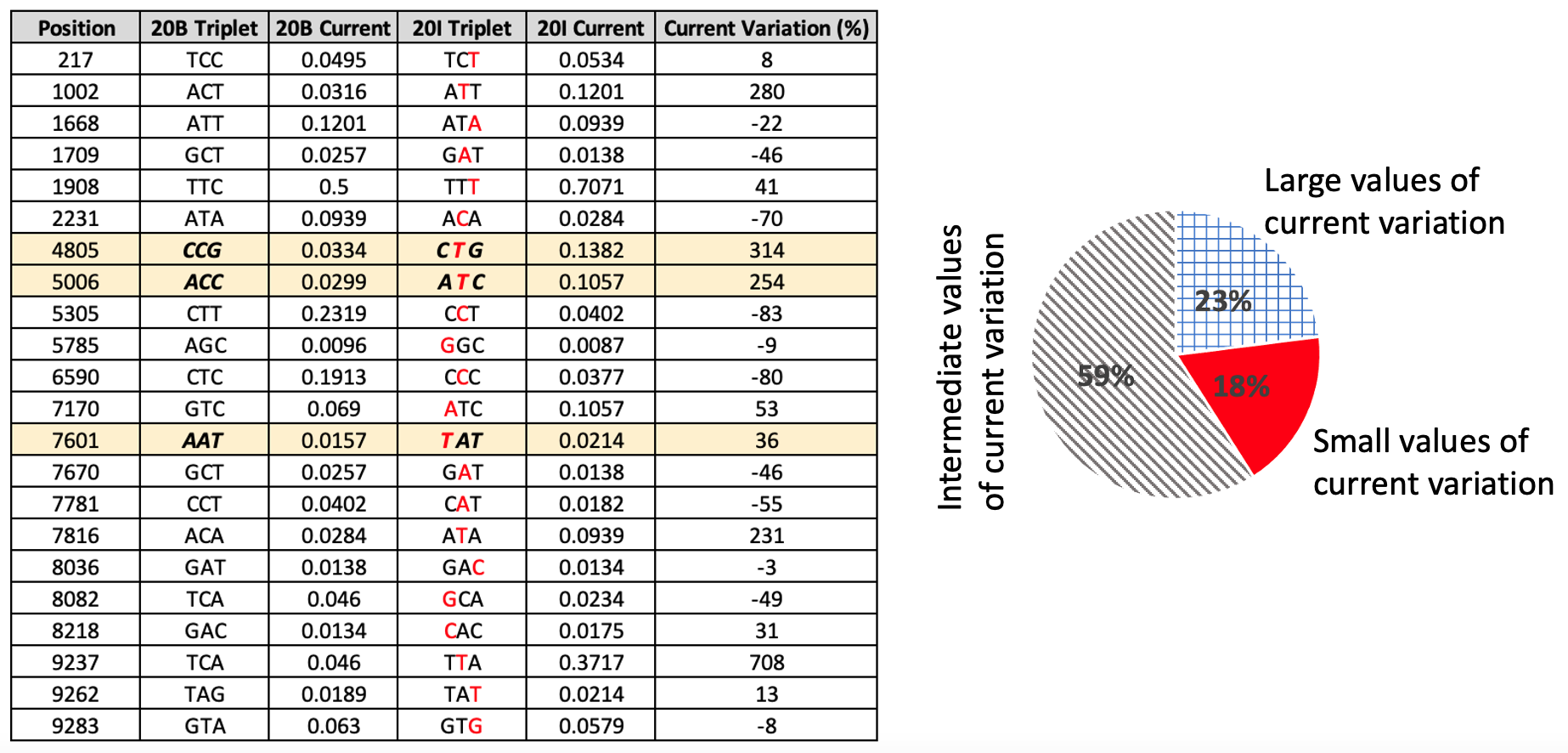}
\\ \textbf{\textbf{Figure 14.}} Comparison between 20B and 20I sequences, with related Chern-Simons current and percentage variation.
\end{center}

\begin{center}
\centering
\includegraphics[width=.91\textwidth]{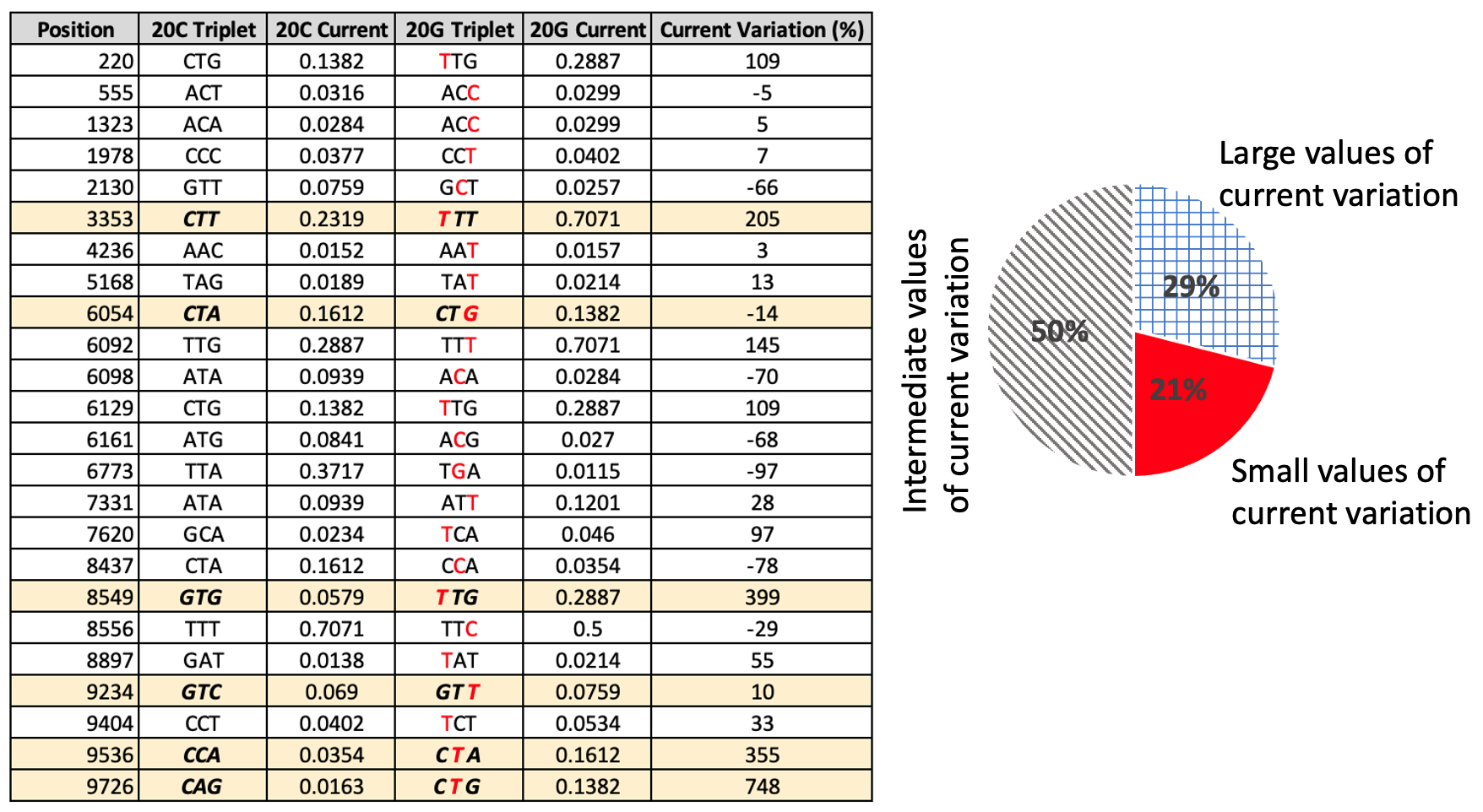}
\\ \textbf{Figure 15.} Comparison between 20C and 20G sequences, with related Chern-Simons current and percentage variation.
\end{center}

\begin{center}
\centering
\includegraphics[width=.91\textwidth]{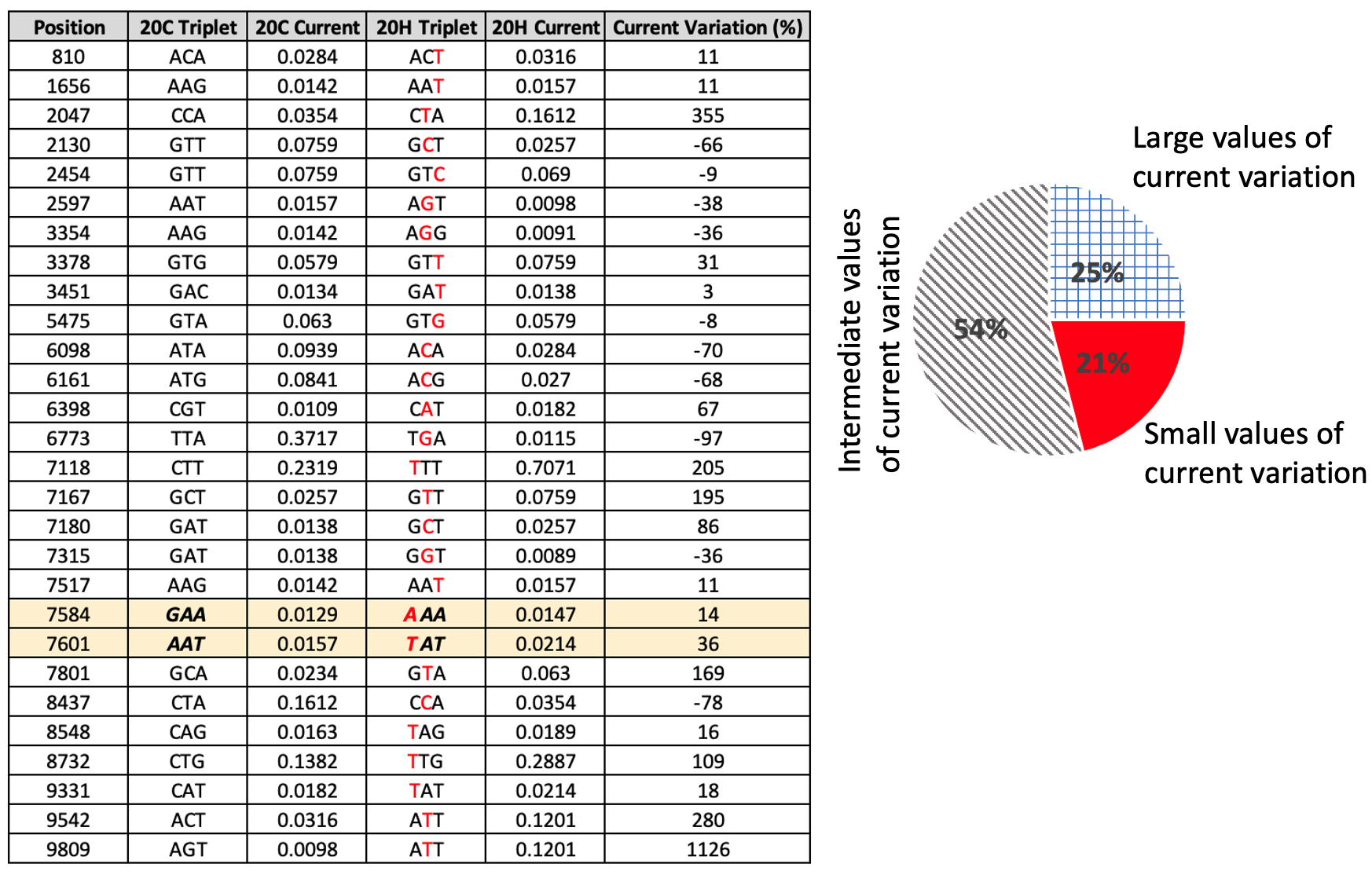}
\\ \textbf{Figure 16.} Comparison between 20C and 20H sequences, with related Chern-Simons current and percentage variation.
\end{center}
\vspace{2 cm}
\begin{center}
\centering
\includegraphics[width=.53\textwidth]{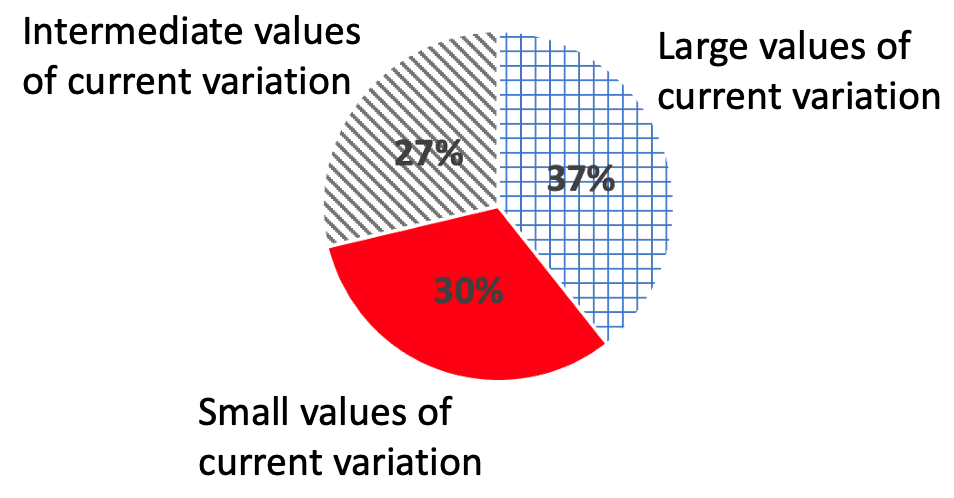}
\\ \textbf{Figure 17.} Details of the whole set of mutations occurring in all sequences.
\end{center}

\newpage

\subsection*{Chern-Simons Current Variations in the Surroundings of Expected Mutations}
\vspace{5mm}
\begin{center}
\centering
\textbf{Table 7.} Chern-Simons currents and their corresponding percentage variations (with respect to the surrounding points) \\ in 19A sequence of SARS-CoV-2 virus. Large values are highlighted in red. 
\includegraphics[width=.60\textwidth]{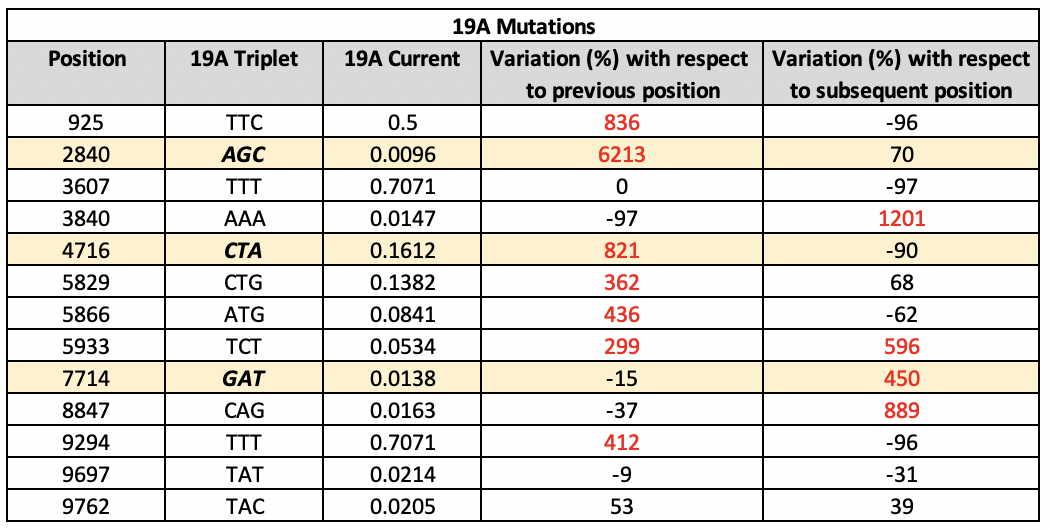}
\end{center}

\begin{center}
\centering
\textbf{Table 8.} Chern-Simons currents and their corresponding percentage variations (with respect to the surrounding points) \\ in 20A sequence of SARS-CoV-2 virus. Large values are highlighted in red. 
\includegraphics[width=.60\textwidth]{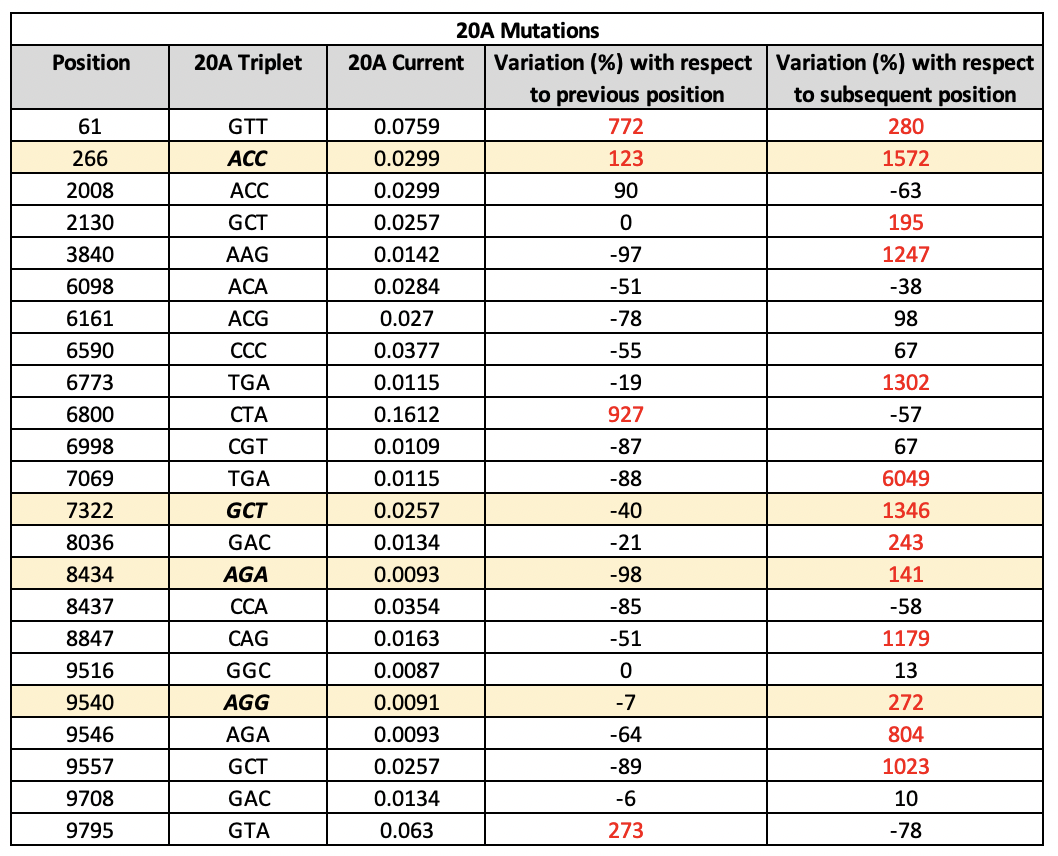}
\end{center}
\newpage
\begin{center}
\centering
\textbf{Table 9.} Chern-Simons currents and their corresponding percentage variations (with respect to the surrounding points) \\ in 20B sequence of SARS-CoV-2 virus. Large values are highlighted in red.
\includegraphics[width=.60\textwidth]{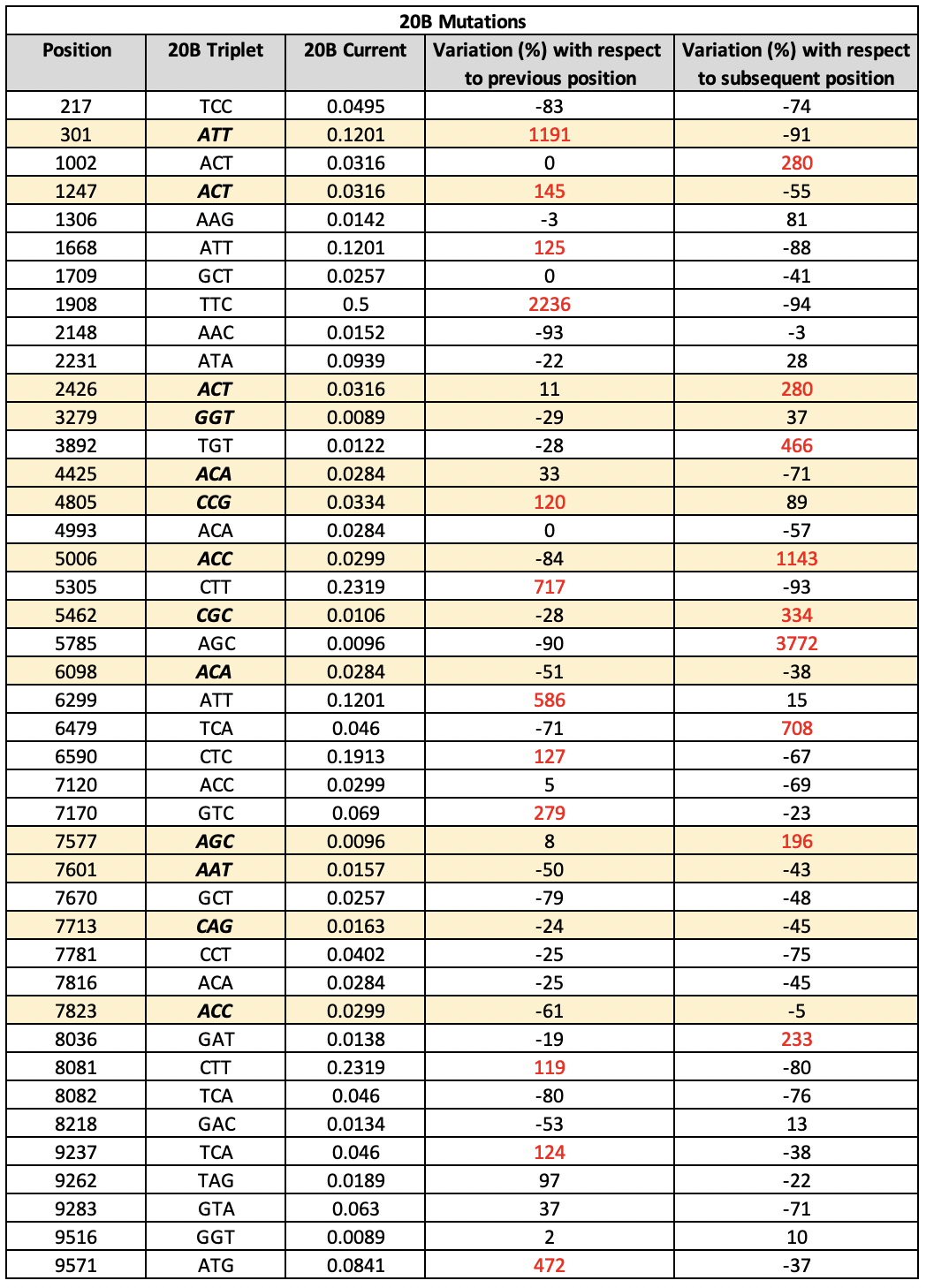}
\end{center}
\newpage
\begin{center}
\centering
\textbf{Table 10.} Chern-Simons currents and their corresponding percentage variations (with respect to the surrounding points) \\ in 20C sequence of SARS-CoV-2 virus. Large values are highlighted in red.
\includegraphics[width=.60\textwidth]{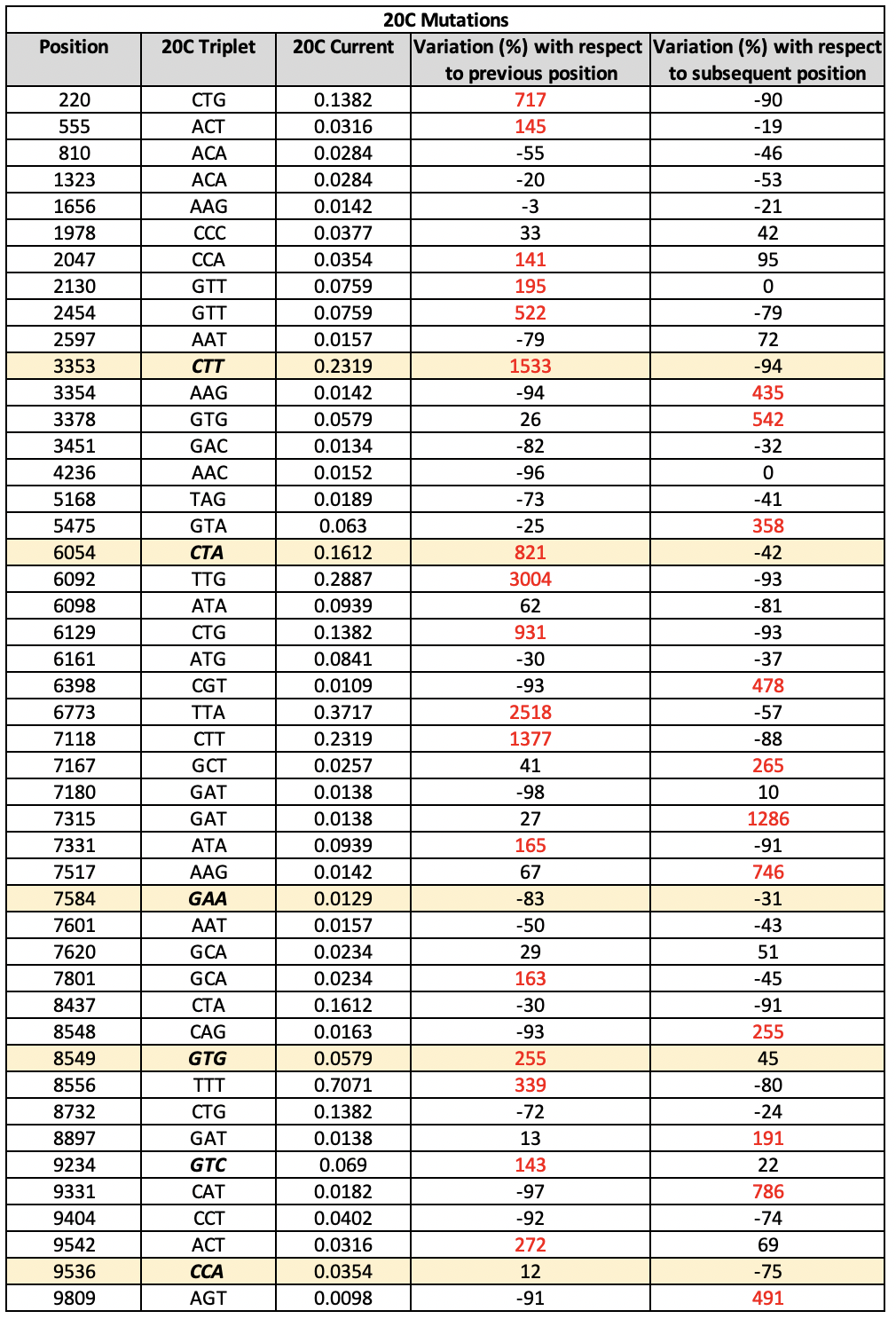}
\end{center}

\begin{center}
\centering
\textbf{Table 11.} List of amino acids of 19A sequence in the spike protein, with corresponding positions, Chern-Simons currents and their variations with respect to surrounding positions. Listed amino acid are those involved in forming the tertiary structure, according to Ref. \cite{Watanabe}. \\
\includegraphics[width=.55\textwidth]{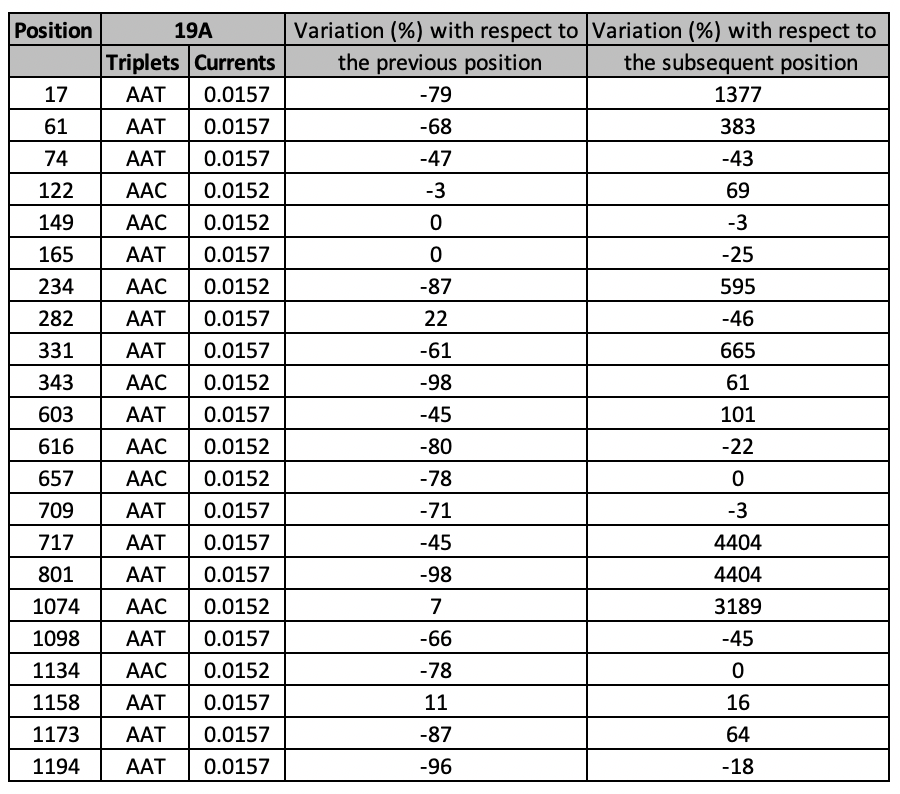}
\end{center}

\end{document}